\documentclass[11pt,a4paper]{article}

\usepackage{jcappub}

\usepackage{natbib}
\usepackage{amsmath}
\usepackage{color}
\usepackage{multirow}
\usepackage{deluxetable}

\tabletypesize{\scriptsize}
\global\let\tablenotemark\relax
\global\def\tablenotemark#1{{\normalfont\textsuperscript{#1}}}
\global\let\tablenotetext\relax
\global\def\tablenotetext#1#2{\tablenotemark{#1}{\scriptsize #2}}

\begin{document}

\graphicspath{{./figs/}}


\def\spider{{\sc Spider} }
\def\spidern{{\sc Spider}}
\def\scip{SCIP }
\def\scipn{SCIP}
\def\fpufs{~$\mathrm{FPU}\!\cdot\!\mathrm{flights}$ }
\def\fpufsn{~$\mathrm{FPU}\!\cdot\!\mathrm{flights}$}
\def\acbar{{ACBAR }}
\def\acbarn{{ACBAR}}
\def\archeops{{\it Archeops}}
\def\cmbpol{CMBpol}
\def\blast{{\sc BLAST}}
\def\scubatwo{{\sc SCUBA-II}}
\def\JIQU{{\sc jiqu}}
\def\ebex{{\sc ebex}}
\def\bicep{{\sc Bicep}}
\def\bicepn{{\sc Bicep}}
\def\biceptwo{{\sc Bicep2}}
\def\boomerang{{\sc Boomerang }}
\def\boomerangn{{\sc Boomerang}}
\def\keck{{\sc Keck}}
\def\QUAD{QUaD}
\def\quiet{{QUIET}}

\def\mytitle{{\Large \spider}: \\
Probing Inflation with a Large Angular Scale Millimeter-wave Polarimeter \\
FY2011 NASA ROSES APRA Suborbital Investigations Proposal \\ PI: William Jones}

\def\emode{$E$-mode}
\def\bmode{$B$-mode}
\def\fsky{$f_{\mathrm{sky}}$}
\def\fskym{f_{\mathrm{sky}}}
\newcommand{\ukrts}{ $\mu\mathrm{K}_{\mathrm{\mbox{\tiny\sc cmb}}}\sqrt{\mathrm{s}}$ }
\newcommand{\ukrtsn}{$\mu\mathrm{K}_{\mathrm{\mbox{\tiny\sc cmb}}}\sqrt{\mathrm{s}}$}
\newcommand{\ukrjrts}{ $\mu\mathrm{K}_{\mathrm{\mbox{\tiny\sc rj}}}\sqrt{\mathrm{s}}$ }
\newcommand{\cl}{$\mathcal{C}_\ell$ }
\newcommand{\cln}{$\mathcal{C}_\ell$}
\newcommand{\te}{$\langle TE\rangle$ }
\newcommand{\ten}{$\langle TE\rangle$}
\newcommand{\ee}{$\langle EE\rangle$ }
\newcommand{\een}{$\langle EE\rangle$}
\newcommand{\boomn}{{\sc Boomerang}}
\newcommand{\boom}{{\sc Boomerang }}
\newcommand{\btk}{{\small B2K} }
\newcommand{\btkn}{{\small B2K}}
\newcommand{\planckn}{{\it Planck}}
\newcommand{\planck}{{\it Planck }}
\newcommand{\planckhfi}{{\it Planck} HFI }
\newcommand{\planckhfin}{{\it Planck} HFI}
\newcommand{\cmb}{{\sc cmb} }
\newcommand{\cmbn}{{\sc cmb}}
\newcommand{\sini}{$\mathrm{Si}_3\mathrm{N}_4$ }
\newcommand{\fwhm}{{\sc fwhm}}
\newcommand{\vsp}{\vspace{3mm}}

\def\bullit{{$\bullet$ }}

\def\nkcmbtxt{{{$\mathrm{nK}_{\mathrm{\mbox{\tiny\sc cmb}}}$}}}
\def\nkrjtxt{{{$\mathrm{nK}_{\mathrm{\mbox{\tiny\sc rj}}}$}}}
\def\mukcmbtxt{{{$\mu\mathrm{K}_{\mathrm{\mbox{\tiny\sc cmb}}}$}}}
\def\mukrjtxt{{{$\mu\mathrm{K}_{\mathrm{\mbox{\tiny\sc rj}}}$}}}
\def\mukcmb{{{\mu\mathrm{K}_{\mathrm{\mbox{\tiny\sc cmb}}}}}}
\def\mukrj{{{\mu\mathrm{K}_{\mathrm{\mbox{\tiny\sc rj}}}}}}
\def\clstar{\ell \left( \ell + 1 \right) C_l / 2 \pi}

\def\mukcmbrts{\mu\mathrm{K}_{\mathrm{\mbox{\tiny\sc cmb}}}\sqrt{\mathrm{s}}}

\def\eg{{{\em e.g.}}}
\def\ie{{{\em i.e.}}}
\def\cf{{{\em c.f.}}}
\def\insitu{{\em in situ}}
\def\etal{{\em et al.}}
\def\rms{{\em r.m.s.}}

\def\mev{MeV~c$^{-2}$}
\def\gev{GeV~c$^{-2}$}
\def\tev{TeV~c$^{-2}$}
\def\cmsq{cm$^2$}
\def\cmcu{cm$^3$}
\def\msq{m$^2$}
\def\mcu{m$^3$}
\def\kmsq{km$^2$}
\def\kmcu{km$^3$}
\def\persec{s$^{-1}$}
\def\peryr{yr$^{-1}$}
\def\perster{ster$^{-1}$}
\def\permsq{m$^{-2}$}
\def\permcu{m$^{-3}$}
\def\percmsq{cm$^{-2}$}
\def\percmcu{cm$^{-3}$}
\def\perkmsq{km$^{-2}$}
\def\perkmcu{km$^{-3}$}
\def\perkev{keV$^{-1}$}
\def\perday{d$^{-1}$}
\def\perkg{kg$^{-1}$}
\def\perdetector{detector$^{-1}$}
\def\gevpercm3{\gev~\percm3}
\def\gpercm3{g~\percm3}
\def\perkmsqperyr{\perkmsq~\peryr}
\def\percmsqpersec{\percmsq~\persec}
\def\kmpersec{km~\persec}
\def\kmpersecpermpc{\kmpersec~Mpc$^{-1}$}
\def\dru{\perkev~\perkg~\perday}
\def\iru{\perkg~\perday}
\def\perdd{\perdetector~\perday}
\def\airu{\percmsq~\perday}
\def\madru{\perm2~\perkev~\perday}
\def\mairu{\perm2~\perday}

\newcommand{\WMAP}{\emph{WMAP}}
\newcommand{\Planck}{\emph{Planck}}
\newcommand{\Archeops}{\textsc{Archeops}}
\newcommand{\ACT}{ACT}
\newcommand{\SPT}{SPT}
\newcommand{\COBE}{\emph{COBE}}
\newcommand{\DASI}{DASI}
\newcommand{\SCIP}{SCIP}
\newcommand{\pending}[1]{\textcolor{red}{#1}}
\newcommand{\sky}{\mathrm{sky}}
\newcommand{\jcap}{J.~Cosmol.~Astropart.~Phys.}
\newcommand{\nar}{New~Astron.~Rev.}


\title{\spidern: Probing the early Universe with a suborbital polarimeter}

\author[1]{A.~A.~Fraisse,}
\author[2]{P.~A.~R.~Ade,}
\author[3]{M.~Amiri,}
\author[4]{S.~J.~Benton,}
\author[5,6]{J.~J.~Bock,}
\author[7]{J.~R.~Bond,}
\author[6]{J.~A.~Bonetti,}
\author[8]{S.~Bryan,}
\author[3]{B.~Burger,}
\author[1]{H.~C.~Chiang,}
\author[9]{C.~N.~Clark,}
\author[9]{C.~R.~Contaldi,}
\author[5,6]{B.~P.~Crill,}
\author[3]{G.~Davis,}
\author[5,6]{O.~Dor\'e,}
\author[7,10]{M.~Farhang,}
\author[5]{J.~P.~Filippini,}
\author[10]{L.~M.~Fissel,}
\author[10]{N.~N.~Gandilo,}
\author[5]{S.~Golwala,}
\author[1]{J.~E.~Gudmundsson,}
\author[3]{M.~Hasselfield,}
\author[11]{G.~Hilton,}
\author[6]{W.~Holmes,}
\author[5]{V.~V.~Hristov,}
\author[11]{K.~Irwin,}
\author[1]{W.~C.~Jones,}
\author[12]{C.~L.~Kuo,}
\author[13]{C.~J.~MacTavish,}
\author[5]{P.~V.~Mason,}
\author[8]{T.~E.~Montroy,}
\author[5]{T.~A.~Morford,}
\author[4,10]{C.~B.~Netterfield,}
\author[9]{D.~T.~O'Dea,}
\author[1]{A.~S.~Rahlin,}
\author[11]{C.~Reintsema,}
\author[8]{J.~E.~Ruhl,}
\author[5]{M.~C.~Runyan,}
\author[5]{M.~A.~Schenker,}
\author[10]{J.~A.~Shariff,}
\author[10]{J.~D.~Soler,}
\author[5]{A.~Trangsrud,}
\author[2]{C.~Tucker,}
\author[5]{R.~S.~Tucker,}
\author[6]{A.~D.~Turner,}
\author[3]{and D.~Wiebe}
\affiliation[1]{\small Department of Physics, Princeton University,
  Princeton, NJ,~USA}
\affiliation[2]{\small School of Physics and Astronomy, Cardiff University,
  Cardiff, UK} 
\affiliation[3]{\small Department of Physics and Astronomy, University of
  British Columbia, Vancouver, BC, Canada}
\affiliation[4]{\small Department of Physics, University of Toronto,
  Toronto, ON,~Canada}  
\affiliation[5]{\small Department of Physics, California Institute of
  Technology, Pasadena, CA, USA} 
\affiliation[6]{\small Jet Propulsion Laboratory, Pasadena, CA, USA}
\affiliation[7]{\small Canadian Institute for Theoretical Astrophysics,
  University of Toronto, Toronto, ON, Canada}
\affiliation[8]{\small Department of Physics, Case Western Reserve
  University, Cleveland, OH, USA}
\affiliation[9]{\small Theoretical Physics, Blackett Laboratory, Imperial 
  College, London, UK}
\affiliation[10]{\small Department of Astronomy and Astrophysics, University
  of Toronto, Toronto, ON, Canada}
\affiliation[11]{\small National Institute of Standards and Technology,
  Boulder, CO, USA} 
\affiliation[12]{\small Department of Physics, Stanford University,
  Stanford, CA, USA} 
\affiliation[13]{\small Kavli Institute for Cosmology, University of
  Cambridge, Cambridge, UK}
\affiliation[]{}
\affiliation[]{E-mail: \href{mailto:afraisse@Princeton.EDU}{afraisse@Princeton.EDU}}

\abstract{%
We evaluate the ability of \spidern, a balloon-borne polarimeter, to
detect a divergence-free polarization pattern ($B$-modes) in the
cosmic mi\-cro\-wave back\-ground~(CMB).  In the in\-flationary scenario, the
amplitude of this signal is proportional to that of the primordial
scalar perturbations through the tensor-to-scalar ratio $r$.  We show
that the expected level of systematic error in the \spider instrument
is significantly below the amplitude of an interesting cosmological
signal with $r=0.03$.  We present a scanning strategy that enables us
to minimize uncertainty in the reconstruction of the Stokes parameters
used to characterize the CMB, while accessing a relatively wide range
of angular scales.
Evaluating the amplitude of the polarized Galactic
emission in the \spider field, we conclude that the polarized emission
from interstellar dust is as bright or brighter than the cosmological
signal at all \spider frequencies (90~GHz, 150~GHz, and 280~GHz), a
situation similar to that found in the ``Southern Hole.''  We show
that two $\sim\!20$-day flights of the \spider instrument can
constrain the amplitude of the $B$-mode signal to $r<0.03$ (99\%~CL)
even when foreground contamination is taken into account.  In the
absence of foregrounds, the same limit can be reached after one 20-day
flight.}


\maketitle


\section{Introduction}
\label{sec:intro}

Over the past decade, our understanding of cosmology has been
revolutionized by ever more precise measurements of the temperature
and of the polarization of the Cosmic Microwave Background (CMB).
The first-year \WMAP~data provided the first full-sky characterization  
of the CMB temperature anisotropies, first detected by
\COBE~\citep{Smoot_etal_1992} and mapped with high fidelity by \boom
\citep{deBernardis_etal_2000}, over a range of angular scales that
allowed for the derivation of stringent cosmological constraints
\citep{Spergel_etal_2003}.  These constraints were 
further strengthened by \WMAP's full-sky measurement of a curl-free
polarization pattern ($E$-mode) in the CMB sky \citep{Page_etal_2007},
a pattern first detected from the ground by
\DASI~\citep{Kovac_etal_2002}.  Many other CMB experiments, both
ground-based and balloon-borne, have provided complementary
measurements, which made it possible to study physics 
poorly constrained by \WMAP~alone.  This is the case of
\QUAD~\citep{quad2009} and \acbar \citep{Reichardt_etal_2009}, whose
data combined with \WMAP's enabled the convincing detection of the
effect of primordial helium on the CMB temperature power spectrum 
\citep{Komatsu_etal_2011}, thereby opening a new window on big bang
nucleosynthesis.  \ACT~\citep{Das_etal_2011} and
\SPT~\citep{Keisler_etal_2011} have provided temperature data down to
arcminute scales, probing the CMB power spectrum up to the
seventh acoustic peak, while \bicep~\citep{bicep_2yr} first detected
the $\ell\sim 100$ acoustic peak in the $E$-mode power spectrum, a
detection later confirmed by \quiet~\citep{quiet2010}.

Remarkably, the six parameters of the simple flat, power-law
$\Lambda$CDM model are sufficient to fit not only all the available
CMB temperature and polarization data, but also a wealth of other
cosmological probes, such as distance measurements from Baryon
Acoustic Oscillations \citep[BAOs,][]{Percival_etal_2010}, and Hubble
constant measurements from Cepheids \citep{Riess_etal_2009}.  Although
it is \emph{possible} to fit more parameters to the data, there is no
indication that any are \emph{needed} \citep{Komatsu_etal_2011}.  The
inflation paradigm, which postulates the existence of a period of
accelerated expansion in the early Universe, is the cornerstone of
this very successful model \citep{Baumann_etal_2009}.  Not only does
it explain the flatness and the homogeneity of the Universe, it also
provides a natural mechanism to generate the primordial Gaussian
density fluctuations that left their imprint in the form of
anisotropies in the CMB temperature and in the distribution of the
large-scale structure we observe today.  However, although inflation
is \emph{consistent} with current observational data, there is no
direct \emph{evidence} to date in favor of this paradigm over its
viable alternatives.  The ekpyrotic scenario
\citep[for a review, see, e.g.,][]{Lehners_2008} is arguably the best
motivated alternative to inflation.  In addition to also being
consistent with current data, it has the added benefit of naturally
avoiding some of the theoretical complications that one encounters
when building an inflationary model.

The predictions of the ekpyrotic scenario and of the inflationary
paradigm differ in one major way.  The latter generically predicts the
existence of gravity wave perturbations in the early Universe, which
would result in an observable divergence-free polarization pattern in
the CMB ($B$-modes), whereas the former excludes the production of
such perturbations at a detectable level \citep{Boyle_etal_2004}.
$B$-modes therefore constitute an unambiguous signature
of inflation.  Moreover, the energy scale of inflation sets the
amplitude of these tensor (gravity wave) perturbations relative to
that of the scalar (density) perturbations in the form of the
tensor-to-scalar ratio $r$.  In addition to validating the
inflationary paradigm, detecting $B$-modes would therefore determine
the energy at which inflation occurs, which
would remove the largest source of uncertainty in inflationary
model~building. 

$B$-modes remain as yet undetected.  \bicep~\citep{bicep_2yr}
derived the strongest constraint on $r$ directly from $B$-mode
nondetection to date: $r<0.72$ (95\% CL).
Using CMB temperature data from \SPT, along with CMB temperature and
polarization data from \WMAP~\citep{Komatsu_etal_2011}, 
distance measurements from BAOs \citep{Percival_etal_2010}, and recent
constraints on the Hubble
constant \citep{Riess_etal_2009}, \citep{Keisler_etal_2011} found an 
upper limit of $r<0.17$~(95\%~CL).  This
constraint is about an order of magnitude above that necessary to rule
out the simplest viable class of inflationary models
\citep{Baumann_etal_2009}.

The \Planck~satellite has already revolutionized our understanding of
the sky at microwave to submillimeter
wavelengths \citep{PlanckReview_2011}.  It has so far 
proven to be a remarkable machine to study the properties
of the interstellar medium, and in particular of the emission from
Galactic dust, which is expected to be the brightest polarized sky
signal in all channels of \Planck's High Frequency Instrument (HFI).
Even though \Planck's sensitivity will be sufficient to
map the diffuse Galactic dust emission and constrain the amplitude of  
its polarized component, it will not be high enough
to produce high signal-to-noise maps of this polarized emission at
scales smaller than $\sim\! 20$ degrees.  This in turn
limits \Planck's ability to probe the cosmological $B$-mode signal
beyond the large-scale reionization bump located below $\ell\sim 10$.
Optimistic forecasts of \Planck's ability to detect or constrain
$B$-modes have been published
\cite[e.g.,][]{Efstathiou+Gratton_2009}.  However, two major
assumptions are made in these studies: (1)~that the \Planck~HFI
is ideal at large angular scales ($1/f$ noise is for example ignored),
and (2) that residuals after foreground removal are small enough that
the induced bias on the determination of $r$ is negligible.  Although
it is difficult to evaluate whether this latter assumption is
reasonable, the former is most likely problematic
(for an early review of the HFI noise, see~\citep{HFI_Instrument_2011}).

\spider \citep{Filippini2010} is a balloon-borne polarimeter designed 
to probe the polarization of the microwave sky with unprecedented
sensitivity and fidelity.  The Antarctic Long Duration stratospheric
Balloon (LDB) platform provides a combination of long flight times,
near-space optical backgrounds, and sensitivity to angular scales that
are inaccessible from even the most favorable ground-based sites.  By
taking advantage of these observing conditions during two flights to
be launched, respectively, in December~2013 and December~2015, \spider 
will produce high signal-to-noise polarized maps of 10\% of the 
sky that are exceptionally clean of polarized Galactic emissions, with the goal
of detecting and characterizing the $\ell\sim 80$ 
acoustic peak in the $B$-mode power spectrum for values of the
tensor-to-scalar ratio $r$ down to $0.03$.
In this paper, we review the current status of the \spider experiment,
and present our observing plans to achieve this science goal.

This paper is organized as follows.  In Section~\ref{sec:instrument},
we briefly describe the \spider~instrument, and evaluate the level of
$B$-mode contamination expected from known sources of 
systematic error.  Section~\ref{sec:polsidelobes} presents a detailed
study of the impact of polarized sidelobes on our observations.  The
expected level of polarized Galactic foregrounds in the \spider field 
is summarized in Section~\ref{sec:foreground_model}, where we also
present a model for the polarized emission from interstellar dust.  In 
Section~\ref{sec:observing}, we provide an overview of our observing
strategy, along with a computation of \spidern's filter transfer
function $F_\ell$.  Section~\ref{sec:bs+fr} details
the reasoning that led to the selection of 90~GHz, 150~GHz, and 
280~GHz as \spidern's observing frequencies, and gives our
expected constraints on $r$ taking into account polarized foreground
contamination.  Our conclusions are summarized in
Section~\ref{sec:final}.


\section{The \spider instrument}
\label{sec:instrument}

\begin{figure*}[t]
\begin{center}
\includegraphics[height=2.2in]{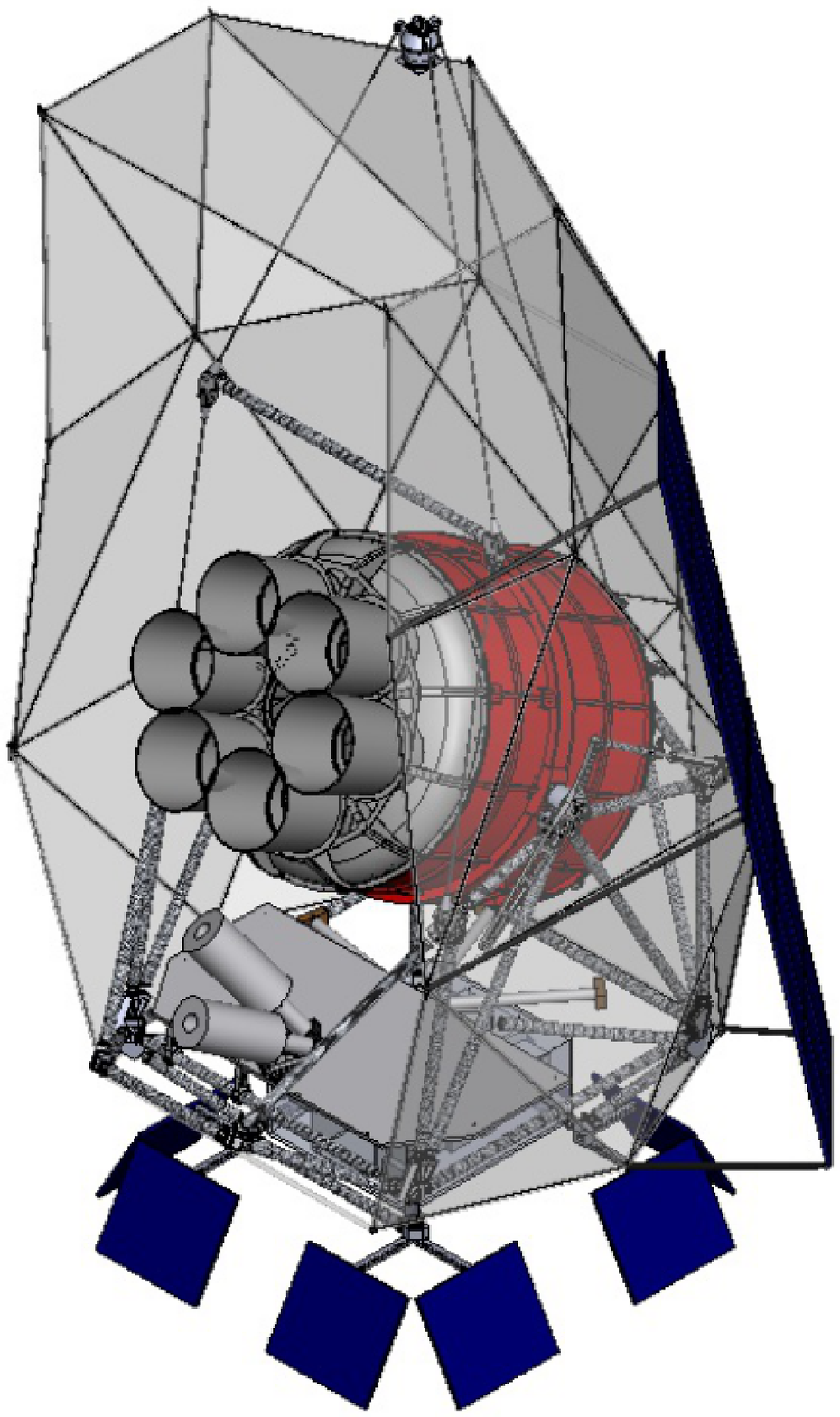}
\includegraphics[height=2in]{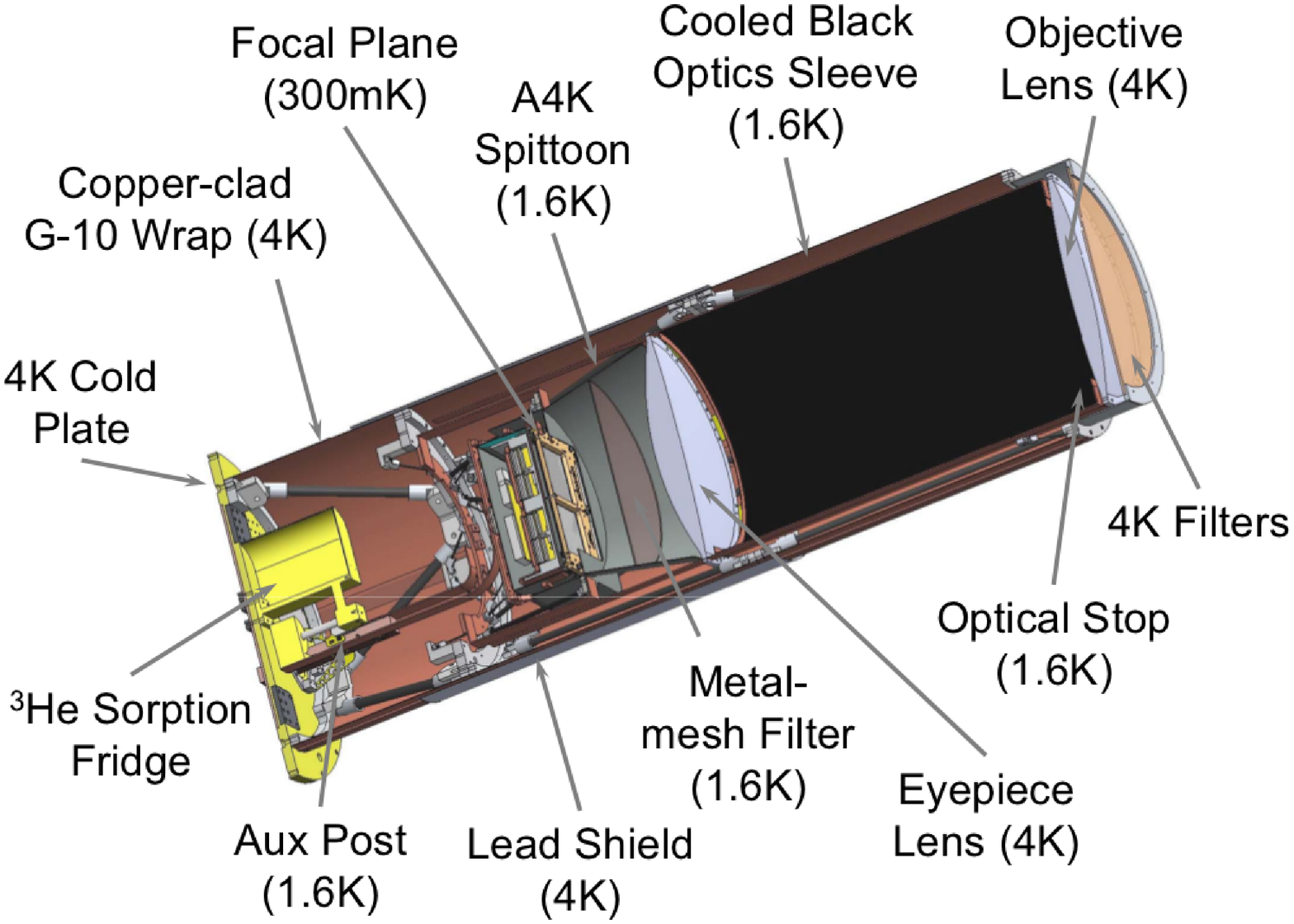}
\includegraphics[height=2.0in]{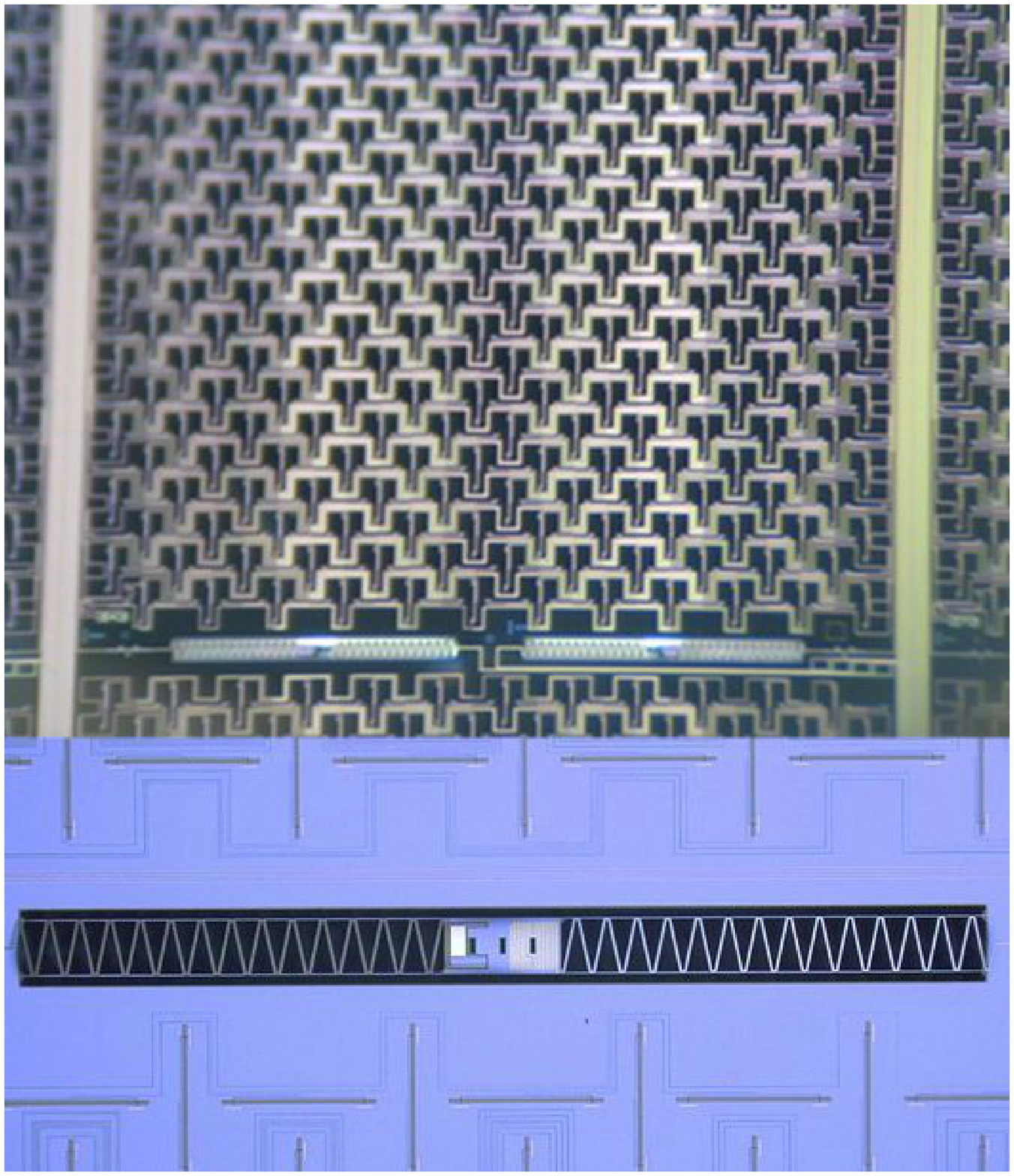}
\end{center}
\caption{{\it Left:} Rendering of the \spider payload, showing the cryostat,
  surrounding frame, and sun shield.  {\it Center:} Cross-section of the \spider telescope
  CAD model with key components labeled. {\it Top right}: Optical pixel,
     showing phased array of slot antennas feeding two TESs at
     bottom. {\it Bottom right}: Close-up of a single TES assembly,
     showing the TES island and the SiN thermal isolation legs.  The
     meandered leg design allows for low thermal conductance in a
     narrow geometry.}
\label{fig:instrument}
\end{figure*}

\subsection{Description}
\label{sec:inst_description}

The \spider payload (Figure~\ref{fig:instrument}) consists of six
mono\-chromatic refracting telescopes derived from the design of the
highly successful \bicep\ instrument.  This compact, axisymmetric
optical system greatly limits the opportunity for polarized
systematics.  The relatively small aperture ($\sim\! 30^\prime$ beam
FWHM at 150~GHz) allows relatively easy baffling of optical sidelobes
while retaining sensitivity to the angular scales relevant to
inflationary science.  The far-field regime of the optics is also
relatively nearby ($\sim\!100$~m), greatly simplifying optical
characterization in the laboratory.

Each refractor illuminates a focal plane of large-format
antenna-coupled transition-edge sensor (TES) arrays
\citep{Kuo:2008spie}.  Each of a focal plane's optical pixels consists
of two interpenetrating arrays of slot antennas sensitive to
perpendicular polarization components.  The power received by each
antenna array is coupled to a dedicated TES through an inline
band-defining microstrip filter.  The antennas, filters and bolometers
for hundreds of polarimetric pixels are fabricated
photolithographically onto a single silicon tile, with no need for
external feed horns or band-defining filters.  The arrays are read out
using a time-domain SQUID multiplexer
system~\citep{dekorte_squid_mux,Battistelli:2008mce}.  Four such focal
planes are currently in operation at the South Pole, as part of the
\biceptwo~instrument \citep{bicep2_spie10} and the Keck array
\citep{keck_spie10}.

\spider will deploy telescopes in three frequency bands centered at
90, 150 and 280~GHz.  Table~\ref{tbl:detectors} lists some of the
major characteristics of a single \spider receiver in each of these
observation bands.  These frequencies are chosen for their sensitivity
to the CMB and to Galactic dust emission.  Since all
frequency-specific optical elements are fully contained within each
telescope insert, \spidern's frequency coverage is easily
adjustable by swapping out one or more of these telescopes.  This
modularity gives \spider a great deal of flexibility in choosing an
optimal frequency coverage for each flight.  Band selection is
discussed further in Section~\ref{sec:bs+fr}.

\begin{deluxetable}{ccccccc}
\tablewidth{0pt}
\tablecolumns{7}
\tablecaption{\spider observing bands, pixel and detector counts, and
  single-detector and single-telescope FPU sensitivities
  \label{tbl:detectors}}
\tablehead{
\colhead{Band Center [Bandwidth]} & \colhead{Beam FWHM} & 
\colhead{Detector} &
\colhead{Detector Sensitivity} & \colhead{FPU Sensitivity} \\
\colhead{(GHz)} & \colhead{(arcmin)} &
\colhead{Count per FPU} &
\colhead{(\ukrts)} & \colhead{(\ukrts)}}
\startdata
90  [22]  &  49  &  288  &  150  &  10 \\
150 [36]  &  30  &  512  &  150  &  7  \\ 
280 [67]  &  17  &  512  &  380  &  18
\enddata
\tablecomments{Each FPU sensitivity is obtained by dividing the
  corresponding single-detector sensitivity by
  $\sqrt{N_\textrm{det}}$, assuming a detector yield of 85\%, slightly
  below the average of the delivered focal planes. The total
  experimental map depth at each frequency scales inversely as the
  square-root of the number of
  $\mathrm{FPU}\!\cdot\!\hspace{0.03cm}\mathrm{flights}$ for that
  frequency. The quoted sensitivities at 90~GHz and 150~GHz are our
  current best estimate based on in-situ measurements of signal and
  noise using an aperture filling 4~K load. The 280~GHz sensitivity is
  scaled from the average in-flight sensitivity of \boom at 245~GHz
  and 345~GHz.  All three bands cover frequency ranges over which
  atmospheric loading, as infered from \citep{Pardo_etal_2001}, is
  minimal.}
\end{deluxetable}

The six receivers
are housed within a shared $\sim\! 1300$~L liquid helium cryostat, with
sub-Kelvin cooling for each focal plane provided by a dedicated $^3$He
sorption refrigerator.  The cryostat is supported below the balloon
within a lightweight carbon-fiber gondola frame, derived from the
\blast~\citep{Pascale2008} design.  A reaction wheel allows the
payload to scan in azimuth, while the elevation of the inner frame is
adjustable with a simple linear actuator similar to the
\boom \citep{Masi2006} design.  Pointing information is provided by
two tracking star cameras, one fixed star camera, rate gyroscopes, 
differential GPS, and a sun sensor.  Extensive sun shielding and
baffling, combined with the relatively small optical apertures of the
\spider telescopes, allows the instrument to scan close to the Sun for
increased sky~coverage.  

The design of \spider has been extensively optimized to take full
advantage of the low mil\-li\-me\-ter-wave backgrounds available from a
stratospheric balloon platform, as well as to ameliorate polarized
systematics to the level needed to characterize \bmode\ polarization.
\spider employs detector arrays very similar to those successfully
fielded in \biceptwo\ and the Keck array, but tuned for much lower
noise-equivalent temperatures.  Extensive filtering and cold ($< 3$~K)
baffling within each instrument greatly reduces stray photon loading
on the bolometers.  \spidern's simple, telecentric optics limit the
contribution of polarized optical sidelobes.  A 4~K half-wave plate
at the aperture of each telescope \citep{bryan_hwp} is stepped in
orientation periodically throughout the observation period in order to
modulate the polarization signal of the sky with respect to any
polarized instrumental systematics.  These systematics have already
been characterized extensively during optical testing of the first
\spider telescopes and through operation of similar technology at the
South Pole.  Tolerances and achieved performance are described in more
detail in Section~\ref{sec:systematics}.
For a more detailed description of the \spider
instrument, we refer the reader
to~\citep{Filippini2010,Runyan2010,Gudmundsson2010}.

\subsection{Systematics review}
\label{sec:systematics}

\begin{deluxetable}{lccl}
\tablecolumns{4}
\tablewidth{0pt}
\tablecaption{Expected sources of systematic error for \spidern
              \label{tbl:Systematics}}
\tablehead{\colhead{Source} & \colhead{Target} & 
  \colhead{$\mathcal{C}_{100}^\mathrm{residual}/\mathcal{C}_{100}^{BB}$} &
  \colhead{Current Status}}
\startdata
Relative gain uncertainty: $\Delta (g_1/g_2)/(g_1/g_2)$ & 0.5\% & 17\% &
  0.1\% Achieved by \boomerangn\tablenotemark{a} \\
Differential pointing: $(\mathbf{r}_1-\mathbf{r}_2)/\sigma$ & 5\% &
  \multirow{3}{2cm}{%
    $\left\}\rule{0mm}{5mm}\right.$\hspace{0.37cm} %
    20\%\tablenotemark{$\dagger$,*}}
    & $2.4\%$ Measured in \spidern \\
Differential beam size: $(\sigma_1-\sigma_2)/\sigma$ & 0.5\% & & 
  0.3\% Measured in \biceptwo\tablenotemark{d} \\
Differential ellipticity: $(e_1-e_2)/2$ & 0.6\% & & 
  $0.15\%$ Measured in \biceptwo \\
Absolute polar angle calibration: $\Delta \Psi_{\mathrm{abs}}$ & $1^{\circ}$ &
  17\% & $0.7^\circ$ Achieved by \bicepn\tablenotemark{b} \\
Relative polar angle knowledge: $\Delta \Psi_{\mathrm{rel}}$ & $1^\circ$ & 6\% & 
  $0.1^\circ$ Achieved by \bicepn\tablenotemark{b} \\
Telescope pointing uncertainty: $\Delta \mathbf{b}$ & $10^\prime$ &
  6\% & $2.4^\prime$ Achieved by \boomerangn\tablenotemark{c} \\
Beam centroid uncertainty: $\Delta \mathbf{c}$ & $1.2^\prime$ &
  12\% & Achieved by \bicep\tablenotemark{b} \\  
Polarized sidelobes (150 GHz): $G_{\mathrm{s}}$ & 
  $-17$ dBi\tablenotemark{$\parallel$} &
  8\%\tablenotemark{$\ddagger$} & Achieved by \bicep\tablenotemark{b} \\
Optical ghosting: $G_{\mathrm{r}}/G_0$ & $-17$ dB & 6\% & 
  Achieved by \biceptwo\tablenotemark{d} \\
HWP differential transmission: $p$ & 0.7\% & 10\%\tablenotemark{$\dagger$} & 
  Achieved by \spidern\tablenotemark{e} \\
Magnetic shielding at focal plane: $\Phi _\mathrm{s}$ & 10 \mukcmbtxt/$B_e$ &
  3\%\tablenotemark{$\dagger$} &
  Achieved by \spidern\tablenotemark{f}
\enddata
\vspace{0.1in}
\raggedright
\tablenotetext{$\parallel$}{Gain at $30^\circ$ from the
  telescope boresight}\\
\tablenotetext{*}{Amplitude of the residual for the three sources of systematic error combined}\\
\tablenotetext{$\dagger$}{\citep{ODea_etal_2011}; 
  $^\ddagger$This paper (see Section~\ref{sec:polsidelobes}); 
  $^{\rm a}$\citep{Masi2006}; 
  $^{\rm b}$\citep{Takahashi2009};
  $^{\rm c}$\citep{Jones_etal_2006};
  $^{\rm d}$\citep{aikin_spie10}; 
  $^{\rm e}$\citep{bryan_hwp}; 
  $^{\rm f}$\citep{Runyan2010}}
\tablecomments{The target for each parameter is set so that the
maximum false \bmode\ signal is less than the \bmode\ power spectrum
at $\ell=100$ for $r=0.03$.  The third column provides the expected
level of residual for each systematic effect if the target is exactly
met \citep[unless otherwise noted, numbers are from][]{mactavish}.
The conventions for beam-related systematic errors follow those in
\cite{bock_epic08}.  $B_e$ is the strength of the Earth's magnetic
field. The optical ghosting simulations of \citep{mactavish} primarily probe
$E \rightarrow B$ contamination and not the potentially more important
$T \rightarrow B$ contribution that would be expected given nontrivial
polarization of the optical ghosts.  We are currently revisiting the
ghosting simulations to incorporate our knowledge of their polarized
response.  Given the frequency dependence of the foreground emission
(see Section~\ref{sec:spider_foregrounds}), requirements for polarized
sidelobes at 150~GHz are sufficient for observations at 90 and
280~GHz.  The \emph{Current Status} column indicates whether each
target is met given current measurements (for \spidern, this
corresponds to lab studies).  When no reference is given, the current
status is based on unpublished measurements.  Adding all systematic
errors in quadrature gives a total false signal of $\sim\!37\%$ of the
$r = 0.03$ \bmode\ signal at $\ell = 100$.}
\end{deluxetable}

Precise control of instrument systematics is crucial for achieving
\spidern's science goals.  Table~\ref{tbl:Systematics} summarizes our
current understanding of a variety of sources of systematic error,
including gain uncertainty, pointing and beam effects, half-wave plate
nonidealities, and sensitivity to the Earth's magnetic field.  We
characterize each systematic effect using a suite of simulations with
no input \bmode\ power, by assuming a target level of control over the
relevant parameters, and measuring the level of the resulting false
\bmode\ signal at $\ell=100$. 
The design requirement is to reduce systematic error to a level of
$(43~{\rm nK})^2$,
the value of the primordial $r=0.03$ \bmode\ power spectrum
at $\ell=100$ in units of $\ell(\ell+1)C_{\ell}/(2\pi)$.

In most cases, the target values for each parameter are derived from a
well-established simulation pipeline described in detail in \citep{mactavish}
and \citep{ODea_etal_2011}.
The simulations include a detailed model of polarized Galactic dust
emission, which interacts nontrivially with instrumental effects. 
The aim of these simulations is to quantify the extent to which a
false \bmode\ signal is produced from systematics that induce
$I \rightarrow Q,U$ or $Q \leftrightarrow U$ mixing, when no attempt is
made to correct for them.   
These simulations assume a scan strategy and observing region that
differ from the current baseline. However, internal simulations done
using the McMurdo observing strategy described in
Section~\ref{sec:obs_overview} suggest that
science requirements will be comparable.  Further simulations are ongoing.

Many of the targets listed in Table \ref{tbl:Systematics} 
for systematics related to design elements common to both \bicep~and
\spider
have been
met by the former \citep{Takahashi2009}, which ensures that 
\spider is in a position to do so as well.
Additional sources of systematic error due to half-wave
plate nonidealities and magnetic field pickup are investigated
in detail in~\cite{ODea_etal_2011}.  In both cases, we find that the
level of \bmode\ contamination is small compared to the $r=0.03$
inflationary signal, even without special effort to correct for these
systematics. 

\spidern's control of systematic errors is further enhanced by the
use, with each telescope,
of a rotating half-wave plate that is
periodically stepped so that the same patch of sky is observed at
multiple orientation angles, thereby augmenting the natural angular modulation
due to the motion of the sky.  This homogenizes the statistical noise
in polarization and provides powerful rejection of a variety of
instrumental systematics.  By locating the half-wave plate skyward of
the telescope aperture, we ensure that only the sky signal is
modulated and that the effects of beam mismatches between detectors
within a pair are greatly reduced. 

Gain drifts contribute significantly to the systematics budget.  \boom
achieved a relative gain uncertainty less than 0.1\%, a level five
times smaller than the target established for \spider
in~\cite{mactavish}, using an on-board calibration lamp
\citep{Masi2006}.  We expect \spider to reach a similar level of
control using regular modulations of the detector bias currents.  This
technique is currently undergoing laboratory tests.  A \boomn-style
calibration lamp is also an option, pending the results of these
studies.

Thermal fluctuations within the instrument will inevitably create a
signal in the \spider bands.  Changes in focal plane temperature
(e.g., with payload altitude) can mimic a sky signal by changing the
power flow through the bolometer support membranes.  \spidern's TES
bolometers self-heat to a temperature largely independent of that of
the focal plane, rendering them less sensitive to such fluctuations
than a comparable semi-conductor bolometer.  Since the monolithic
fabrication of each detector array ensures good uniformity in thermal
response within each polarization pair, such fluctuations do not
propagate as strongly into a false polarization signal.  A passive
thermal filter, consisting of stainless steel heat capacity blocks
between the detectors and thermal strap, supresses any thermal
fluctuations on time scales comparable to the scan (by a factor of 
$\sim\! 20$~dB at 30~mHz).

Since about 25\% of each beam terminates on the optical stop, changes
in the stop temperature will also mimic a sky signal.  Essentially all
such internal sidelobes terminate on the cooled ($\sim\! 1.8$~K)
optics sleeve, however, whose low temperature limits the sensitivity
of the bolometers to such drifts.

We report on simulations specifically designed to
evaluate the level of the contamination expected from polarized
sidelobes in the next section.

\subsection{Polarized sidelobes}
\label{sec:polsidelobes}

\begin{figure*}[t]
\centering
\includegraphics[width=2.6in]{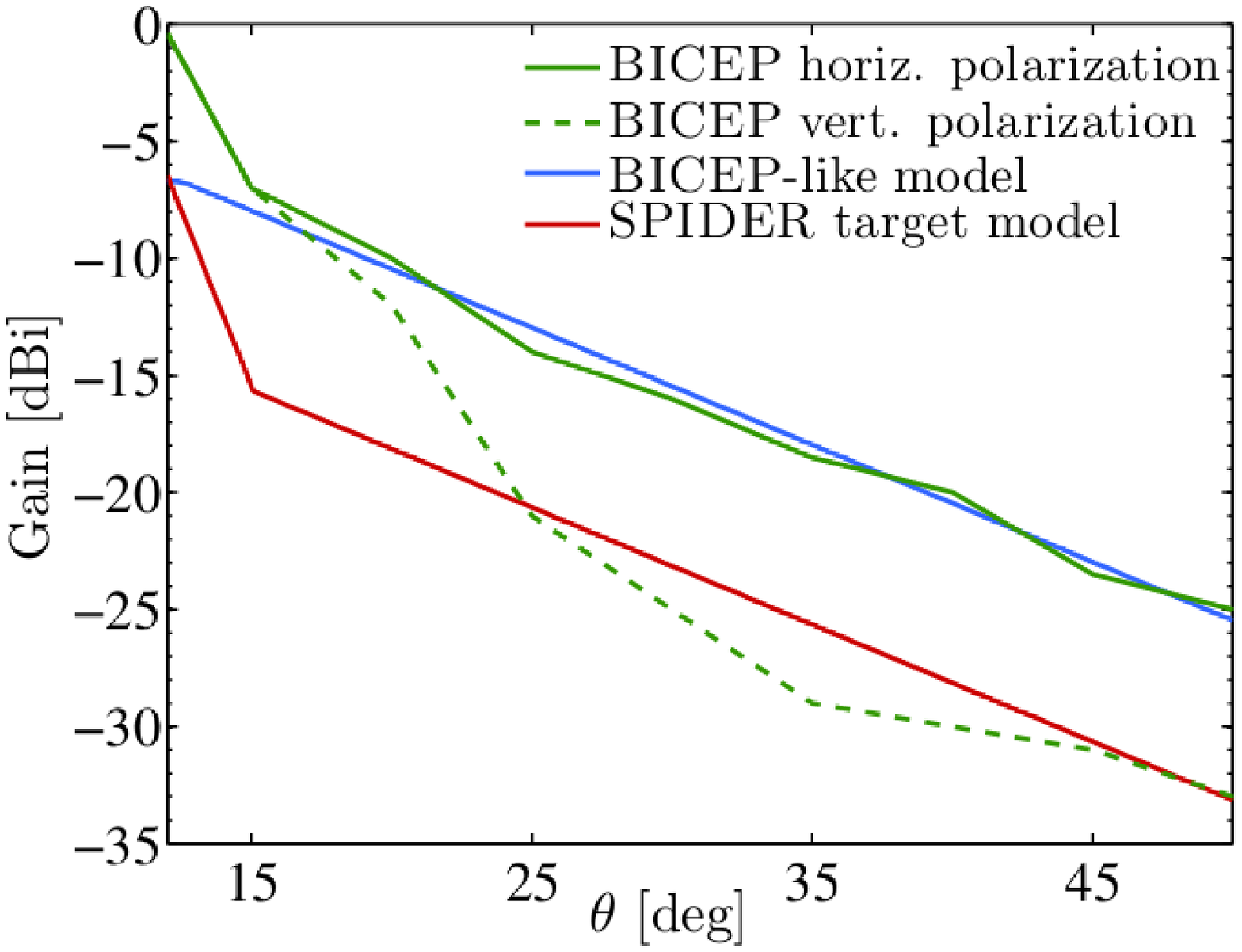}\hspace{5mm}
\includegraphics[width=2.6in]{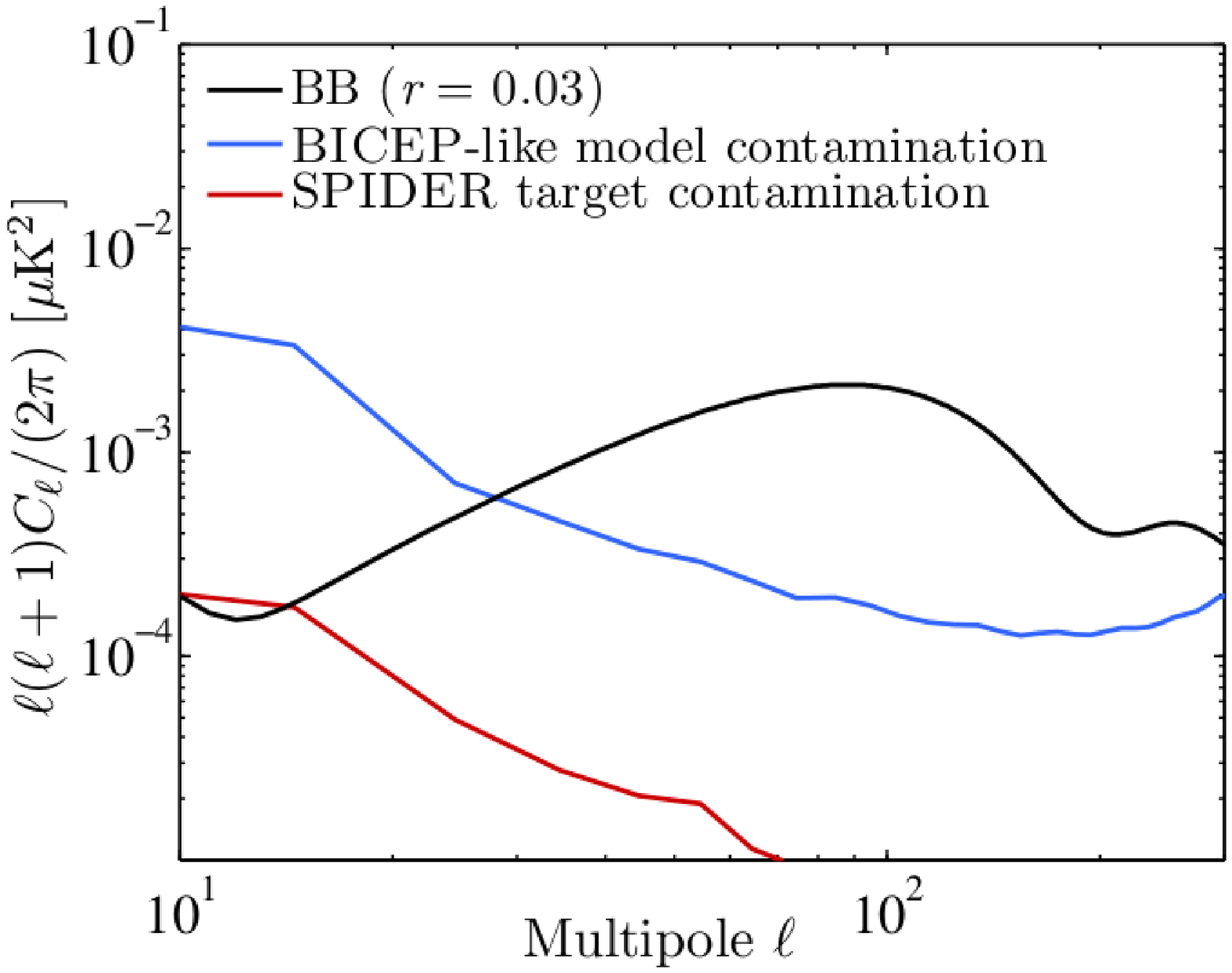}
\caption{{\it Left:} Polarized sidelobe profiles measured at 150~GHz in
  \bicep\ (green) and models used in the analysis presented in
  Section~\ref{sec:polsidelobes}.  We consider two models: a sidelobe
  profile comparable to \bicep's (blue), and \spidern's target profile,
  which assumes an additional 10~dB attenuation starting at $12^\circ$
  from the telescope boresight (red).  {\it Right:} \bmode\ power
  spectra of the sidelobe pickup obtained with the sidelobe
  profiles shown in the left panel.  The $r=0.03$ primordial $B$-mode
  power spectrum is shown for comparison.  The contamination is well
  below the cosmological signal at $\ell = 100$ for both profiles.
  However, the \bicep-like model leads to substantial contamination at
  large angular scales~($\ell\lesssim 25$).  \spidern's target profile
  ensures that the false $B$-mode signal is comparable to or
  fainter than the cosmological signal at low multipoles, and
  negligible at $\ell\sim 80$, where the primordial $B$-mode power
  spectrum~peaks.
\label{fig:polsidelobe}}
\end{figure*}

Coupling through the beam sidelobes to emission from the Sun, the
Earth, and the Galactic plane will produce a spurious polarized
signal.  In order to establish mission requirements for \spidern's
sidelobe response, we convolve a polarized sidelobe profile with a
temperature map of the Galactic emission at 150~GHz based on the sky
model developed in \citep{ODea_etal_2011} and discussed further in
Section~\ref{sec:dust_model}.  Because of its complex morphology and
proximity to \spidern's observing region, the Galaxy is indeed
expected to represent the most significant source of contamination on
the scales of interest.  \spidern's optical system is a cryogenic
refracting telescope, based on that developed for \bicep. Unlike
\bicep, \spider employs a conical baffle with its base located inside
the cryostat and extending roughly $0.5\:\mathrm{m}$ out from the
surface of the vacuum vessel.
 The rim
of the baffle is asymmetric to prevent direct illumination of the
half-wave plate by the balloon and the Earth's limb, and intercepts
the field of view between $10^\circ$--$12^\circ$, in contrast with the $15^\circ$
cutoff angle for the \bicep\ telescope forebaffle.  \spidern's main
beam and near ($2^\circ$--$12^\circ$) sidelobe profiles are informed
by a physical-optics model of the ideal optical chain internal to the
cryostat, which includes both lenses and the half-wave plate.
Accurately modeling \spidern's far~($>\!12^\circ$) sidelobes requires
a complete simulation of the optical system including sun shield,
baffle, gondola, and balloon that has yet to be performed.  Instead,
we first model \spidern's far sidelobe profile as a power law out to
$50^\circ$, where the beam is truncated.  This model is consistent
with measurements of \bicepn's polarized sidelobe response
\citep{Takahashi2009}, which is representative of \bicepn\rq s
relatively less restrictive forebaffle. The profile we use is shown in the
left panel of Figure~\ref{fig:polsidelobe}.

The \bmode\ power spectrum derived from the sidelobe model described
above is shown in the right panel of Figure~\ref{fig:polsidelobe}.
Compared to the primordial $r=0.03$ \bmode\ signal, it is over
12~times fainter at $\ell=100$, while it is significantly brighter at
large angular scales~($\ell\lesssim 25$).

Unsurprisingly, the \bicep~forebaffle, which was designed to provide
rejection at the level of $r\sim 0.1$ while observing at relatively
high Galactic latitudes, does not provide sufficient rejection of the
foreground signal for \spidern\rq s large-scale measurements.  By
adding $\sim\! 10$~dB of attenuation beyond $12^\circ$ off axis as
shown in the left panel of Figure~\ref{fig:polsidelobe}, the expected
sidelobe signal decreases to a level near or below the $r=0.03$
$B$-mode spectrum at all scales of interest.  The contamination is
then over an order of magnitude fainter than the cosmological signal
at $\ell\sim 30$, and entirely negligible at the $\ell\sim 80$ peak
(see Figure~\ref{fig:polsidelobe}).  Although further testing is
required to demonstrate that \spidern\rq s baffle will provide the
required factor of ten improvement over the \bicep~measurement, it
appears to be well within reach.  For instance, a simple diffraction
calculation shows that adding a one-inch diameter diffraction edge on
the sun shield at the edge of the baffle aperture leads to a gain at
$12^\circ$ off axis that is $\sim\! 20$~dB lower than in the
BICEP-like model.  This additional attenuation remains above 10~dB all
the way to $50^\circ$ off axis.


\section{Galactic foregrounds}
\label{sec:foreground_model}

\subsection{Polarized foregrounds in the \spider field}
\label{sec:spider_foregrounds}

During its first two flights, \spider will map the microwave sky at
frequencies ranging from 90~GHz to 280~GHz.  At 90~GHz,
the large-scale ($\ell \lesssim 10$) polarized emission from
interstellar dust is expected to be at least an order of magnitude
brighter than the $r=0.03$ primordial $B$-mode signal that \spider
aims to detect~\citep{Dunkley_etal_2009}.  At smaller scales,
the angular power spectrum of the polarized dust emission appears to
be compatible with the power-law dependence
$\mathcal{C}_\ell^\mathrm{dust} \propto \ell^{-2.6}$
\citep{Cho+Lazarian_2010}, which results in 
its amplitude being comparable to that of our target $B$-mode signal
at $\ell \sim 40$.  Models indicate that this multipole dependence might
break down in favor of a steeper decline at multipoles higher than
$\sim\!10^3$.  Assuming that the polarized emission from interstellar
dust obeys the same frequency dependence as its total emission, these
results can be extrapolated to higher frequencies
using the best-fit FDS model~\citep{Finkbeiner:1999p19}.
The resulting angular power spectra are $\sim\!9$ and
$\sim\!5\times 10^2$ times higher than the 90~GHz signal at 150~GHz and
280~GHz,~respectively.  Throughout this paper, we assume spinning dust
emission to be unpolarized, consistent with theoretical expectations 
\citep[see, e.g.,][]{Lazarian+Finkbeiner_2003}.

Among other Galactic foregrounds, only synchrotron emission is
expected to be significantly polarized at microwave frequencies.
We use the \WMAP~23~GHz data to evaluate the amplitude of the
polarized Galactic synchrotron emission in the \spider region at
frequencies ranging from 90~GHz to 280~GHz.  The extrapolated signal
is negligible both at 150~GHz and at 280~GHz compared to the $r=0.03$
$B$-mode signal.  At 90~GHz, its large-scale power spectrum is 
a factor of two higher than that of the target
signal, but a factor of five lower than that of the polarized emission
from interstellar dust.  We find the multipole dependence of the
polarized synchrotron emission to be well described by the power law
$\mathcal{C}_\ell^\mathrm{synchrotron} \propto \ell^{-2.5}$
\citep[compatible with \WMAP's full-sky estimate in][]{Page_etal_2007},
which brings this spectrum under that of the $r=0.03$ $B$-mode
signal by $\ell\sim\!30$.  At the $\ell\sim 80$ $B$-mode peak, this
foreground is an order of magnitude fainter than the
target signal.  As a result, we do not expect the Galactic
synchrotron emission to limit \spidern's ability to detect $B$-modes
at the $r=0.03$ level.

Since the Galactic plane will be masked during the cosmological
analysis, we do not consider the polarization of the free-free
emission that might be induced by Thomson scattering at the edges of
H\textsc{ii} clouds \citep{Rybicki+Lightman_1979}.  Free-free emission
is otherwise unpolarized.

Given the considerations in this section, we expect the polarized
emission from interstellar dust to be the one Galactic foreground that
will hinder \spidern's ability to detect a primordial $B$-mode
signal.  We show in Section~\ref{sec:dust_model} that it is
possible to select observing fields accessible from a McMurdo flight
in which the polarized emission from interstellar dust is up to an
order of magnitude fainter than discussed above at all \spider
frequencies.  However, even with this reduced level of contamination,
Galactic dust is still the dominant polarized sky signal at and above 
90~GHz.  Section~\ref{sec:bs+fr} therefore addresses the
question of how Galactic emission and cosmological signal can be
separated given an appropriate combination of frequency bands.
We describe the Galactic foreground model required for both these
analyses in Section~\ref{sec:dust_model}.

\subsection{Model for the polarized emission from interstellar~dust}
\label{sec:dust_model}

A model of the sky that \spider will observe was presented in detail
in \citep{ODea_etal_2011}.  In this section, we summarize the
procedure followed to create a template of the polarized
emission from interstellar dust that takes into account the
three-dimensional Galactic structure.

The nonsphericity of interstellar dust grains and their ability to
align with the Galactic magnetic field (hereafter, the Field) were
first put forth in 
an effort to explain the polarization of starlight observed
independently by Hiltner and by Hall \citep{Hiltner_1949,Hall_1949}.
With its long axis preferentially aligned perpendicular to the
Field \citep[see, e.g.,][]{Hoang+Lazarian_2008}, a dust grain absorbs
more radiation in the direction
perpendicular to the local Field line than in the direction parallel
to it.  This results in a net linear polarization of the incident
starlight in the direction parallel to the Field.  The polarization of
the emission from dust grains is the counterpart of this absorption
phenomenon.  As a result, the emission from
interstellar dust is expected to be polarized 
perpendicularly to
the sky-projected direction of the Field.
\Archeops~and \WMAP~both
detected the polarization of the Galactic dust emission at
microwave frequencies
with high significance \citep{Benoit_etal_2004,Page_etal_2007}, and
found degrees of polarization over the sky compatible with theoretical
expectations \citep[][]{Draine+Fraisse_2009}.  However, neither
experiment had the combination of resolution and sensitivity necessary
to produce a map of this emission usable by \spidern.  As a result,
we must rely on modeling to estimate the characteristics of the
polarized emission from interstellar dust
in the \spider field.

Despite tremendous recent progress in our understanding of the
alignment of interstellar dust with the Galactic magnetic field
\citep[see, e.g., the review in][]{Draine_2011}, we are still far from
a full theoretical understanding of this physical process.
Modeling the polarized emission from Galactic dust therefore requires
simplifying assumptions.  It is customary to assume (i) that each
grain is aligned with the Field with its long axis exactly (as opposed
to preferentially) perpendicular to the field line; (ii) that, given a
field strength, each grain has the same polarizing effect; and (iii)
that the degree of polarization of the dust emission induced by a
grain is proportional to the square of the magnetic field strength at
the location of the grain.  With these
assumptions, and given a three-dimensional model of the dust
distribution and of the Galactic magnetic field, we compute a
full-sky template of the polarization fraction and angle of the
Galactic dust emission.  Details of the three-dimensional models and
of the calculation performed to derive the template are
provided in \citep{ODea_etal_2011}.  Since the magnetic field model
includes a small-scale turbulent component, the resulting
depolarization effect is taken into account.  The overall amplitude of
the template for the polarized fraction of the Galactic dust emission
is a free parameter, which we set to match the average value of
$3.6\%$ derived by \WMAP~outside of the Galactic plane
\citep{Kogut_etal_2007}.  We obtain polarized intensity maps by
multiplying the template by the dust intensity map from
\citep{Finkbeiner:1999p19}, using their model 8 to account for its
frequency dependence.

\begin{figure*}[t]
\centering
\includegraphics[width=2.5in]{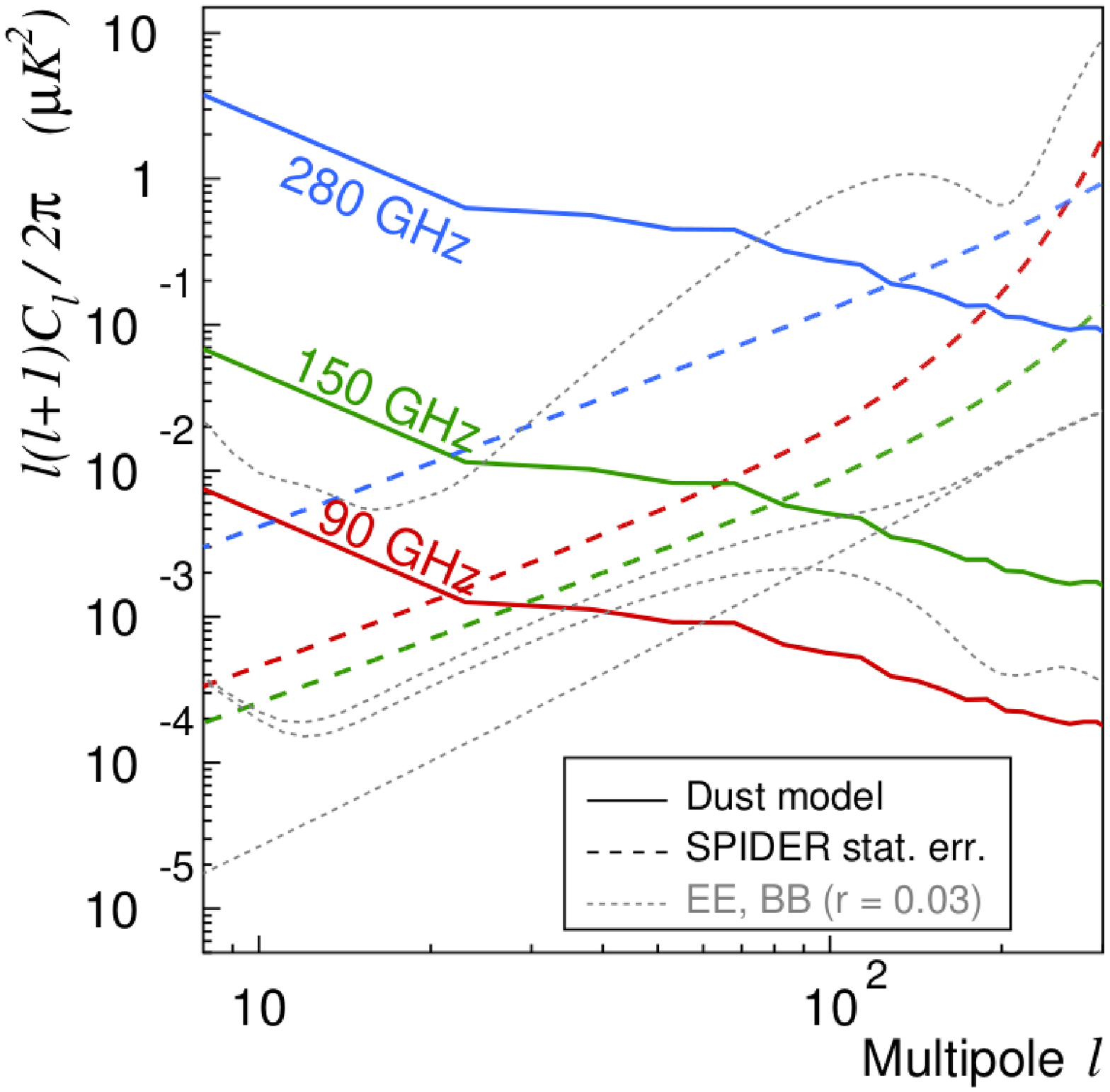}\hspace{5mm}
\includegraphics[width=2.5in]{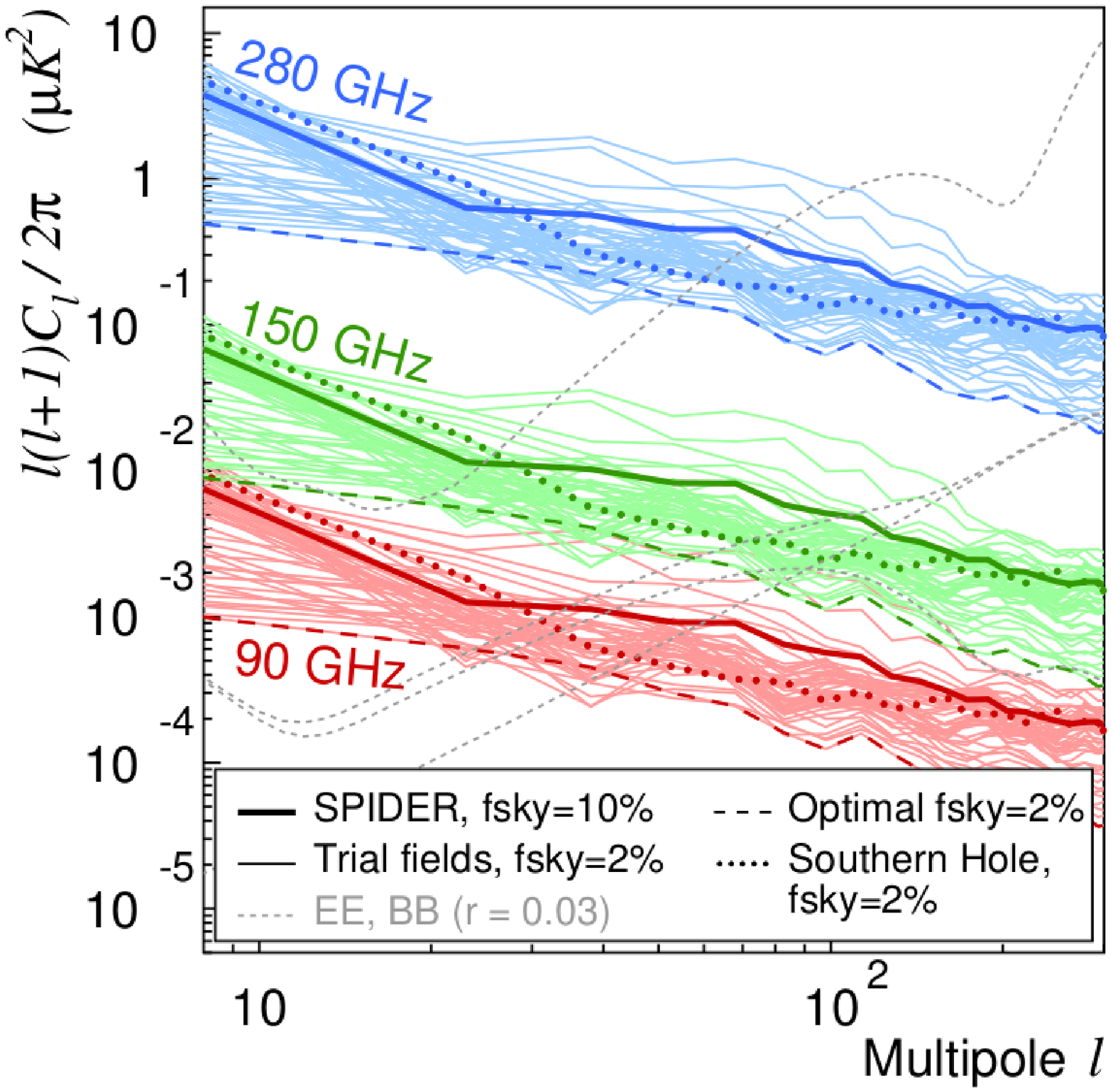}
\caption{{\it Left:} A multipole-by-multipole comparison of the
  levels of statistical noise at each frequency in
  Table~\ref{tbl:detectors} after both 
  flights and of the astrophysical and cosmological signals,
  including Galactic foregrounds, lensing, and the B-mode signal
  assuming a scalar-to-tensor ratio $r=0.03$.
  {\it Right:}~Level of Galactic foreground emission in the
  nominal \spider field ($f_\sky=10\%$) compared to constituent trial
  fields with $f_\sky=2\%$.  The optimal 2\% of the sky, outlined in
  Figure~\ref{fig:coverage}, has the lowest polarized dust emission at
  $50 < \ell < 200$ not only within the \spider region, but also
  within the entire area of sky accessible from a McMurdo LDB flight.
  Spectra are also shown for a trial field centered on
  ($\alpha=0^\circ$, $\delta=-57^\circ$), which lies within the
  ``Southern Hole,'' a target region used by 
  several ground-based experiments. Note that our foreground model
  does not include polarized synchrotron emission from the Galaxy
  given the weakness of this signal in our bands in the \spider region 
  (see discussion in Section~\ref{sec:spider_foregrounds}).
  \label{fig:fgcompare}}  
\end{figure*}

The left panel of Figure~\ref{fig:fgcompare} shows the power spectra
of the polarized Galactic dust emission in the \spider field for all
three frequencies in Table~\ref{tbl:detectors}, along with the
per-multipole statistical error for each band after both flights as
predicted by a Fisher analysis (see Section~\ref{sec:bs+fr}).  At a
given frequency, the brightness 
of the polarized emission from interstellar dust in the \spider field
is comparable to that of the full-sky average of this signal (Galactic
plane excluded), whose characteristics were described in 
Section~\ref{sec:spider_foregrounds}.  Interestingly, it is also
comparable to that of the polarized Galactic dust emission in a
typical $f_\sky=2\%$ patch in the popular ``Southern Hole'' region, as
shown in the right panel of Figure~\ref{fig:fgcompare}, even though
\spider will observe five times as much sky as covered by this patch.
This appears to indicate that although the Southern Hole is believed
to be the region of the southern sky most free of dust emission,
it might not be the region most free of \emph{polarized} dust emission.
Finally, it is worth noting that the \spider region encompasses the
cleanest 2\% of the sky accessible from a McMurdo LDB flight, and that
a large majority of its component fields exhibit significantly less polarized
Galactic dust emission than the region average (the relevant power
spectra are also shown in the right panel of
Figure~\ref{fig:fgcompare}).  This will provide valuable cross-checks
to evaluate the level of foreground contamination in the
\spidern~maps.


\section{Observing strategy}
\label{sec:observing}

\subsection{Overview}\label{sec:obs_overview}


\begin{deluxetable}{ll}
\tablecolumns{2}
\tablewidth{0pt}
\tablecaption{\label{tab:flightinfo}}
\tablehead{\multicolumn{2}{c}{Summary of general information on the \spider mission}}
\startdata
Launch Location & McMurdo Station, Antarctica \\
Launch Date & 12/2013 (Flight 1), 12/2015 (Flight 2) \\
Flight Duration (target) & 20 days per flight \\
Altitude (target) & 36,000 m \\
Flight Path & Circumpolar, typically %
              $73^\circ {\rm S} < {\rm latitude} < 82^\circ {\rm S}$ \\
Sky Coverage & $f_\sky\sim 0.1$, $10\lesssim \ell\lesssim 300$
\enddata
\tablecomments{The flight schedule provided in this table is consistent
  with the state of hardware development as of August 2012.}
\end{deluxetable}


\begin{figure*}[t]
\centering
\includegraphics[width=0.9\textwidth]{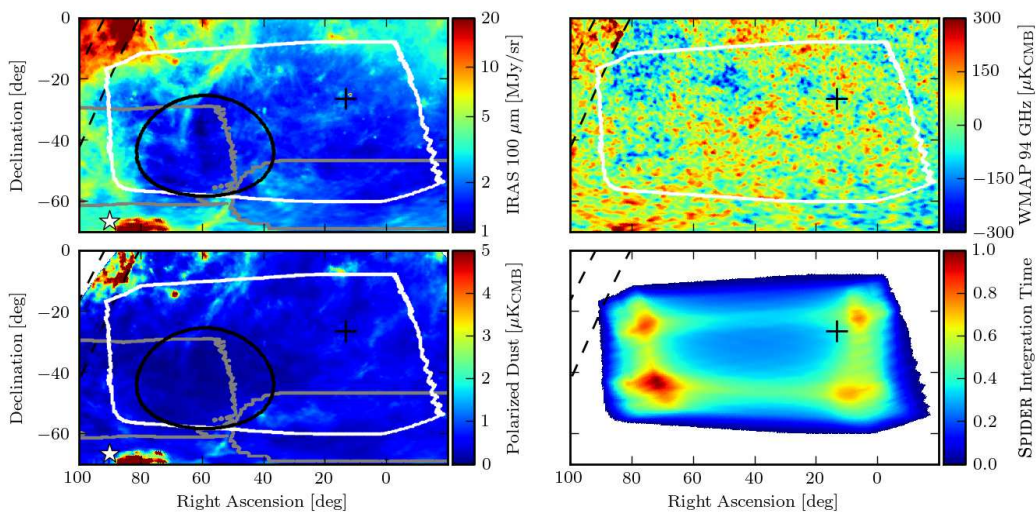}
\caption{\spider observing region and scan profile.  All four panels show
  the same portion of the sky, in equatorial coordinates, smoothed
  with a 30' beam.  The southern Galactic pole (black $+$) is overplotted,
  along with the 10- and 20-degree Galactic latitude lines~(dashed).
  {\it Top~left:}~IRAS 100~$\mu$m data showing
  dust morphology, the \spider observing region (white outline), and
  the south
  ecliptic pole (white~$\star$).  Also shown are the \boom and \bicep\
  fields (left and right gray outlines, respectively),
  and the region of minimum foreground contamination
  in the \spider field (black outline).  {\it Bottom
  left:} Polarized dust emission amplitude at 150~GHz, according to the model in
  Section~\ref{sec:dust_model}.  Clear differences in
  dust emission morphology are visible between this model and the IRAS data.
  {\it Top~right:}~\WMAP~94~GHz $TT$ data in the same area, showing
  relative absence of foreground contamination in 
  the \spider observation region.
  {\it Bottom right:} Distribution of
  integration time, averaged over all detectors in a single
  150~GHz focal plane, for the observing strategy in Section 
  \ref{sec:obs_overview}.
  \label{fig:coverage}}
\end{figure*}

\spider will launch its first flight from McMurdo Station in
Antarctica in December 2013, and observe the southern sky for about 20
days.  The flight parameters are summarized in
Table~\ref{tab:flightinfo}.  The observing strategy is designed to
maximize 
sky coverage and crosslinking over the cleanest regions of the sky
accessible from a McMurdo LDB flight.  Figure~\ref{fig:coverage} shows
the intensity and polarization of the Galactic dust emission in the
\spider region \citep{ODea_etal_2011}, as well as the relatively
foreground-free CMB sky in the 94 GHz \WMAP~band.

The baseline scan strategy consists of azimuthal scans punctuated by
elevation steps and rotations of the half-wave plate.  The scan
velocity profile is sinusoidal with a maximum acceleration of 0.8
degrees/s$^2$ (for a scan period of $\sim\! 45$ seconds).  The exact
scan amplitude and center is adapted at each elevation step to (i)
avoid observing within 90 degrees of the Sun, as set by the design of
the sun shield, and (ii) maintain the instrument boresight pointing at
the region of the southern sky outside of the Galactic plane (0 to 80
degrees right ascension).  Elevation steps occur every hour, in sync
with local sidereal time (LST).  We reach maximum elevation (40
degrees) at 12:00~LST, and minimum elevation (28~degrees) at
24:00~LST.  By syncing the elevation steps with sidereal time, we
ensure maximal sky coverage in declination.  The sky coverage repeats
each sidereal day, providing both redundancy and a rich set of
consistency tests.  Once per day, when sky rotation is at a minimum at
local midnight or noon (depending on when observations begin), we
rotate the half-wave plate of each instrument by 22.5$^\circ$, thereby
switching $Q$ to $U$ in the instrument~frame.

We set aside time after each half-wave plate rotation to perform beam
and pointing calibrations on Galactic sources.  Every half hour, near
the beginning and end of each constant-elevation period, we perform a
brief ($\sim\! 5$ s) in-flight measurement of detector responsivity by
slightly stepping the detector biases. 
Each sub-Kelvin cooler
must be re-cycled every 72 hours. To reduce the instantaneous load on
the 1.6~K stage, these cycles must be staggered.  The cycling of one
instrument has no impact on the others, allowing for a $\sim\! 85$\% 
observational duty cycle.  The process of cycling all six coolers
requires four hours for all instruments to return to the nominal base
temperature.

We estimate the effect that the observing profile will have on the
Stokes $T/Q/U$ reconstruction, assuming an ideal case of no beam
asymmetry, no $T\rightarrow Q,U$ leakage (i.e.~perfect polarization
efficiency), and no noise correlation between detectors.  We calculate
the pointing matrix $A_p$ for each pixel $p$, as
\begin{equation}
A_p = \left<\left[ \begin{array}{ccc}
1 & \cos 2\psi & \sin 2\psi\\
\cos 2\psi & \cos^2 2\psi & \cos 2\psi\, \sin 2\psi \\
\sin 2\psi & \cos 2\psi\, \sin 2\psi & \sin^2 2\psi
\end{array} \right]\right>_p,
\end{equation}
where $\left<\ldots\right>_p$ indicates the average over all
$i=1\ldots N_{{\rm obs},p}$ observations along $p$ at the series of polarization angles
$\{\psi_{p,i}\}$.
The inverse of $A_p$ gives an estimate of the signal $T/Q/U$
covariance for each pixel.  In the ideal case, we would have
$A_p^{-1} = \mathrm{diag}(1,2,2)$.  We therefore construct a figure of
merit, which we call the ``fractional excess variance in polarization,''
from the $QQ$ and $UU$ elements of $A_p^{-1}$:
\begin{eqnarray}
F_p & = & \frac{1}{2}
\left[\sqrt{\left(A^{-1}_p\right)_{QQ}}-\sqrt{2}\right]^2 \nonumber
\\ & & +\, \frac{1}{2}
\left[\sqrt{\left(A^{-1}_p\right)_{UU}}-\sqrt{2}\right]^2.
\end{eqnarray}
\begin{figure*}[t]
\centering
\includegraphics[width=0.9\textwidth]{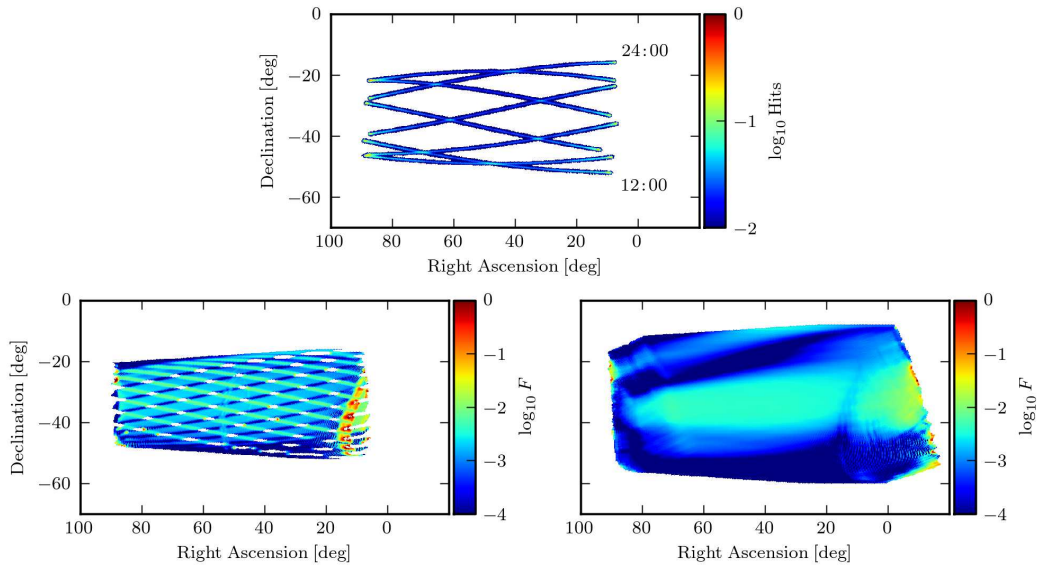}
\caption{The \spider scan profile.  All three panels show the same
  portion of the southern sky, in equatorial coordinates, smoothed
  with a 30' beam.  {\it Top:} A 24-hour period for a single detector,
  illustrating the change in telescope orientation throughout the
  day. Five-minute periods every three hours are shown.  Detector
  pairs in a single pixel are oriented 22.5 and 112.5 degrees relative
  to the scan direction.  {\it Bottom left:} Fraction of excess
  variance (see definition in Section~\ref{sec:obs_overview}) due to
  anisotropic angular coverage in the Stokes $Q$ and $U$ maps for the
  above detector over a four-day period.  {\it Bottom right:} Fraction
  of excess variance for a full focal plane of 512 bolometers, over a
  four-day period.  This observing profile covers 9.5\% of the sky, of
  which 96\% is observed with near-isotropic coverage in crossing
  angles to reconstruct the Stokes $Q$ and $U$ parameters with $<1$\%
  excess variance.\label{fig:crosslink}}
\end{figure*}
This figure of merit is shown in Figure~\ref{fig:crosslink}, for a
single detector, and for a full focal plane 
of 256 pixels (2 detectors per pixel).  The excess variance is below
$1$\% over 96\% of the observed region.  We will fly with pairs of
telescopes clocked at 45$^\circ$ relative to each other, which allows
a simultaneous measurement of both $Q$ and $U$ parameters,
significantly improving crosslinking in the maps and further reducing
the excess variance.

\subsection{Filter transfer function $F_\ell$}
\label{sec:fell}

The interplay between scan strategy and instrumental and
environmental effects plays a significant role in all CMB
experiments.  These instrumental effects include scan-synchronous artifacts,
ground-fixed signals, atmospheric contamination, and low-frequency instability
in the instrument.  Signal estimation pipelines tend to be tailored to
each experiment (or flight of the same experiment) according to the
particular combination of effects that are most relevant to each dataset.

While these algorithms, which include template removal, common mode
decorrelation, and (statistical) least-squares mapmaking, are generally
quite effective at removing systematic effects from the data, they
nevertheless result in a degradation of sensitivity and the loss of
fidelity to particular spatial modes of the astrophysical signal
relative to an idealized dataset consisting only of the signal and
Gaussian uncorrelated white noise.  \spider is no exception, and we
study the impact of these effects through time domain simulations of the
experiment that capture as fully as possible the range of
instrumental and environmental effects that are present~in~the~data.

The impact on the fidelity of the recovered CMB power spectrum can be
approximately characterized via a calculation of the transfer function
$F_\ell$, as determined by the comparison of a
modest ensemble of signal simulations to the input sky \citep{master}.
An understanding of this transfer function is necessary to relate
NET estimates of any experiment, for example those in Table~\ref{tbl:detectors},
to errors on CMB power spectra or experimental upper limits.
As we will show, the relatively benign stratospheric balloon
environment coupled with \spidern\rq s large sky coverage provide
polarized maps of the sky with extremely low noise and high fidelity
to angular scales ranging from the scale of the map to that of the beam.

For \spidern, we expect the large angular scales to be dominated by noise at
time scales below the half-period of an azimuthal scan, which varies
between $\sim\!20$--$25$~s, depending on the scan amplitude.
Contributions to the noise include a scan synchronous component
from the residual atmospheric column as well as $1/f$ noise from
instability on the bias and instrumental backgrounds.  While the
stationary Gaussian component of the noise can be optimally accounted for by
the use of a least-squares mapmaker \citep[see, e.g.,][]{psb_methods},
we expect, based on our experience with past balloon experiments, the
need to apply additional filtering to remove a scan-synchronous
component and reduce the impact of uncertainty in the low-frequency
transfer function.

We measure the net effect of signal processing on the
power spectrum by calculating the transfer function $F_\ell$
following the formalism developed in \cite{master}.  We calculate the
power spectrum $\hat C_\ell^{\rm (in)}$ of a signal-only CMB realization
that has been smoothed by the beam ($30^\prime$ FWHM for the 150~GHz
instrument).  We then apply the observing profile shown in
Figures~\ref{fig:coverage} and~\ref{fig:crosslink} to generate
noise-free 
time-ordered data for processing. These simulated data are then input
to the signal estimation pipeline, which incorporates our estimate of
the noise spectra, flagging and any additional filtering. We then
calculate the power spectrum $\hat C_\ell^{\rm (out)}$ of the resultant maps.
Both the input and output maps are weighted by the
distribution of integration time shown in Figure \ref{fig:coverage} to
ensure identical cut-sky treatment.  The transfer function is then
simply the ratio of the input and output power spectra, averaged over
an ensemble of realizations.

\begin{figure}[t]
\centering
\includegraphics[width=0.46\textwidth]{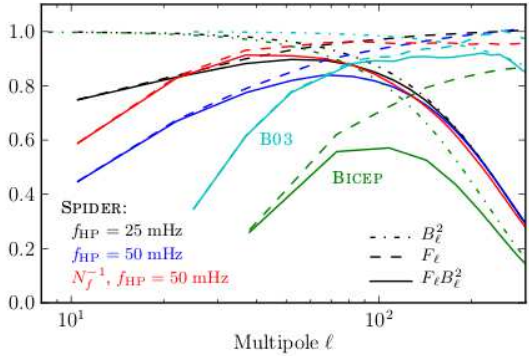}
\caption{An estimate of the filter transfer function $F^{TT}_\ell$ and
  beam function $B_\ell$ for \spidern's 150~GHz instrument, shown
  alongside transfer functions for typical ground-based
  \citep[\bicep,][]{bicep_2yr} and balloon-borne
  \citep[\boomn03,][]{Jones_etal_2006_TT} experiments.  The transfer
  function for \spider is estimated from an ensemble of signal
  realizations assuming time-domain filtering of signals below the
  azimuthal scan frequency.  The black and blue curves were calculated
  from high-pass filtered signals naively binned into maps, while the
  red curve was obtained using an iterative least-squares mapmaker
  that took into account both a high-pass filter and the expected
  bolometer noise PSD.  The low-frequency noise is well approximated
  by a power-law with a spectral index of $-0.75$ and a knee at
  0.15~Hz.
  \label{fig:fell}}
\end{figure}

Figure \ref{fig:fell} shows the published transfer functions for
\bicep\ and {\boomn}03, as well as two estimates for \spider that
correspond to optimistic and pessimistic levels of scan-synchronous
noise.  Results are shown for the temperature anisotropy power
spectrum only; measurements of the polarized filter function are
ongoing, but are not qualitatively different.  We multiply the
transfer function with the beam to illustrate the combined effect on
the power spectrum.  At large angular scales, we lose modes due to
interplay between the filtering and the scan, while small angular
scales are downweighted by the beam.  The relatively severe required
filtering applied in \bicep\ reduces the product of the transfer
function and beam window function to nearly 50\% at the peak.  In the
optimistic scenario for \spidern, we set the high-pass filter cutoff
just above the azimuthal scan frequency (25~mHz), which leads to the
peak nearing 90\% at $\ell\sim50$ even with a naively binned map.
More realistically, we have set the cutoff near 50~mHz, corresponding
to a half-period of the scan.  This choice is informed by the data
from \boomn, and corresponds to the most stringent filtering that had
to be applied to the polarization data from the 2003 flight (towards
the end of the flight, at the relatively low altitude of $\sim\!
30$~km).  \spider preserves fidelity to larger scales (lower
multipoles) than \boom simply because of the larger sky coverage and
higher scan rate, and not because the \spider data are expected to be
qualitatively more stable or well behaved than the \boom data.  While
this more aggressive filtering reduces the transfer function at
$\ell\sim10$ by over 50\%, the peak at $\ell\sim70$ remains relatively
high at 85\%.

The covariance of the band power estimates derived from the data
depends on the effect of the experiment on both sample and statistical
variance.  The sample error on the measured power spectrum $\hat
C_\ell$ increases as $1/\sqrt{F_\ell}$ as the effective number of
observed modes is reduced.  While this increase in the sample variance
has no impact on the ability of an experiment to set upper limits on
the $B$-mode spectrum, the impact of the processing on the statistical
variance does.  

To account for detector noise, we use an iterative least-squares
mapmaker similar to that in \citep{psb_methods} on the filtered data.
The inverse-noise kernel is estimated from a typical noise power
spectrum, measured in our system with no optical loading on the
bolometers.  The absence of (Gaussian and white) photon noise in this
measurement provides an accurate estimate of the instrumental noise.
The significant features in the noise power spectrum are (1) a $1/f$
knee at $\sim\! 100$~mHz, and (2)~excess noise at frequencies above
$\sim\!10$~Hz, likely due to the thermal architecture of the bolometer
island, which reduces the asymptotic transfer function by $\sim\!
5\%$.  The iterative method improves the net transfer function by up
to 30\% relative to the naively binned case at the largest scales.
Recent bolometer noise measurements under optical loading with
improved bias stability show that the $1/f$ knee is likely to be
closer to 25~mHz in flight, which should further improve the
large-scale transfer function.


\section{Band selection \& prospects for foreground~removal}
\label{sec:bs+fr}

In this section, we evaluate the ability of \spider to detect
inflationary $B$-modes as a
function of the frequency distribution of
$\mathrm{FPU}\!\cdot\!\hspace{0.03cm}\mathrm{flights}$ (number of
focal plane units multiplied by number of flights).
In addition to
the 90~GHz, 150~GHz, and 280~GHz channels whose characteristics are
given in Table~\ref{tbl:detectors}, we consider the possibility of
adding a 220~GHz frequency band.  This frequency was chosen to be half
way between \spidern's 150~GHz and 280~GHz channels, thereby
eliminating the frequency coverage gap that results from the frequency
distribution in Table~\ref{tbl:detectors}.  The 220~GHz band would
have the same number of
detectors per FPU as the 280~GHz band.  Its bandwidth and its beam's
FWHM would be 53~GHz and 21$^\prime$, respectively.  The sensitivity
of each of its detectors would be
320~$\mu$K$_\mathrm{\textsc{cmb}}\sqrt{\mathrm{s}}$,
for an overall FPU sensitivity of
15~$\mu$K$_\mathrm{\textsc{cmb}}\sqrt{\mathrm{s}}$.
For programmatic reasons, we require that observations at each
frequency during each flight be performed by at least two FPUs.  
For each \spider flight,
we therefore have the option to acquire
data at up to three frequencies to be chosen among 90~GHz, 150~GHz, 220~GHz,
and 280~GHz.  Since, as discussed in
Section~\ref{sec:foreground_model}, the emission from Galactic dust is
expected to 
become stronger with increasing
frequency, Galactic foregrounds are expected to be lower at 90 and 150~GHz
than at 220 and 280~GHz.  As a result, we choose 90~GHz and 150~GHz as two
of the three \spider bands for the first flight, and perform
numerical simulations to distribute
the remaining two FPUs.

We simulate a hundred maps at each frequency with independent CMB and noise
realizations.  CMB maps are obtained with \WMAP's best fit $\Lambda$CDM
cosmological parameters, $r=0.03$, and CMB lensing included.  We approximate
the noise in \spider as white, governed by the parameters in
Table~\ref{tbl:detectors} (for the 220~GHz band, we use the parameters
listed in the previous paragraph), and weighted by the
expected number of observations
per pixel for the observing strategy in
Section~\ref{sec:obs_overview}.  Each simulated map 
is added to a template for the Galactic dust emission at the appropriate
wavelength.  Dust templates are extrapolated from the
90~GHz map from \citep{ODea_etal_2011} described in
Section~\ref{sec:foreground_model} assuming a single power-law dependence on
frequency with spectral index $\beta_\mathrm{d}=1.7$, the latter being
the average of the one-index best fit to the Finkbeiner-Davis-Schlegel
(FDS) model 8 
\citep{Finkbeiner:1999p19} and of \WMAP's ``base" fit dust spectral
index
\citep{Gold_etal_2009}.  Note that we do not attempt to use a
two-component dust model.  In some of the cases we consider in this
section, this would indeed lead to more parameters than we can possibly
fit given the number of observing bands in play.

With two pairs of FPUs occupied by either 90~GHz or 150~GHz detectors,
we use a simple pixel-by-pixel least-square foreground fitting procedure to
evaluate \textsc{Spider}'s first flight's ability to detect $B$-modes
when the two remaining
FPUs are assigned to (i)~90~GHz and 150~GHz, (ii)~220~GHz, or
(iii)~280~GHz.  In
order to have similar signal-to-noise ratios at the two low-foreground
frequencies, we do not consider cases in which the numbers of FPUs at
90 and 150~GHz differ.  We model the data as
\begin{equation}
\label{eq:fgfit}
\mathbf{S} = \mathbf{S}_\mathrm{CMB} + \mathbf{S}_\mathrm{d,\nu_0} %
             \left(\frac{\nu}{\nu_0}\right)^{\beta_\mathrm{d}},
\end{equation}
where $\mathbf{S}$ is the usual set of Stokes parameters, the index d refers
to dust, and $\nu_0 = 90$~GHz,
and fit the seven free parameters to the simulated data.  In all
cases, we include as part of the data a \Planck~217~GHz map simulated
in the same way as the \spider maps
with the instrument characteristics published in the \Planck~early papers
\citep{HFI_Instrument_2011}.

Given \spidern's
focus on characterizing the $\ell\!\sim\!80$ $B$-mode peak, we use the
$\ell\!=\!80$ $B$-mode signal-to-noise ratio (SNR) computed from
residual maps 
as a figure of merit to
assess the quality of the CMB reconstruction.
Among the three cases we
consider for the first \spider flight, having three FPUs at
90~GHz and three FPUs at 150~GHz
leads to the highest $B$-mode SNR at $\ell\!=\!80$, and therefore to
the best reconstructed CMB map for our purposes.
This configuration turns out to be also favored
from a detector development point of view.

\begin{figure}[t]
\centering
\vspace{0.1cm}
\includegraphics[width=0.45\textwidth]{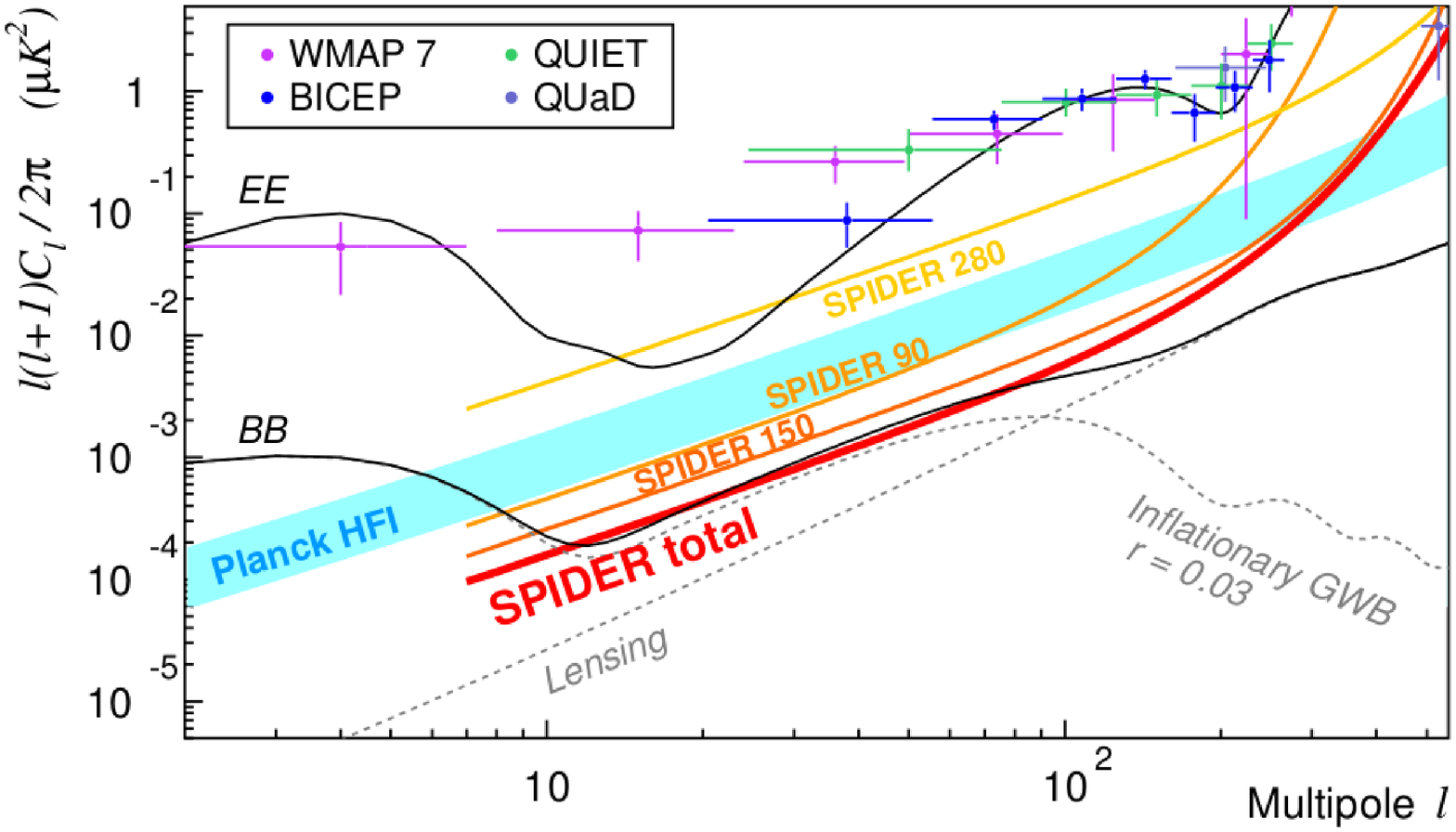}
\caption{Theoretical \emode\ and \bmode\ ($EE$ and $BB$) power spectra
  and projected per-multipole statistical errors for \spider after
  both flights and for
  \planckhfin.  \emode\ detections from \WMAP~\citep{wmap7_params},
  \bicep~\citep{bicep_2yr}, \quiet~\citep{quiet2010}, and
  \QUAD~\citep{quad2009} are also shown for comparison.  The
  \bmode\ spectrum is the sum of two components (dotted lines):
  inflationary gravitational waves, shown here for $r=0.03$, and weak
  gravitational lensing of \emode\ polarization.  The noise curves for 
  \spider and \planck are derived from a simple Fisher analysis,
  assuming no foreground contamination.
  It is likely that foreground emission will limit \planckn's
  measurement of the $BB$ bump below $\ell \simeq 10$ (corresponding
  to an angular scale of $\gtrsim 20$ degrees) induced by
  reionization.  \spider is optimized to cover the $\ell \sim 80$ peak 
  in the gravitational wave spectrum.  The \spider band powers have
  been truncated at a multipole of seven as the limited sky coverage
  will prevent us from probing larger scales.
  \label{fig:cl_spectra}}
\end{figure}

After the first \spider flight, we expect the per-multipole
statistical error at 150~GHz to be at a level comparable to that of the power
spectrum of our foreground model at $\ell\!\sim\!80$.
With Galactic dust
now limiting our
ability to detect inflationary 
$B$-modes, we seek to increase \spidern's frequency coverage in order 
to gather multi-frequency foreground information, which will help
constrain the 
model in Equation~(\ref{eq:fgfit}).
Given the programmatic constraints previously mentioned and continued
observations at 90~GHz and 150~GHz, this can be
achieved only by flying a pair of either 220~GHz or 280~GHz FPUs.
Performing an analysis similar to that described above for the
first \spider
flight, we select the 280~GHz band over the 220~GHz channel, which
leads to the $\mathrm{FPU}\!\cdot\!\hspace{0.03cm}\mathrm{flight}$
distribution listed in 
Table~\ref{tab:fpuflights} along with the cumulative noise in each
band.  Figure~\ref{fig:cl_spectra} shows the resulting per-multipole
statistical error in each band after both flights, as well as the
overall statistical error for the full mission, both derived from a
Fisher analysis.  The expected \Planck~HFI statistical error from a
similar analysis is also shown for comparison.

\begin{deluxetable}{ccccc}
\tablecolumns{5}
\tablewidth{0pt}
\tablecaption{\spider FPU frequency distribution and per-band
  cumulative noise \label{tab:fpuflights}}
\tablehead{\colhead{Flight} & \colhead{FPU Distribution} &
  \multicolumn{3}{c}{Cumulative Noise %
  ($\mu\mathrm{K}_\mathrm{\textsc{cmb}}/\mathrm{deg}^2$)} \\
 & & \colhead{90~GHz} & \colhead{150~GHz} & \colhead{280~GHz}}
\startdata
\spider 1 & $3 \times 90$~GHz; $3 \times 150$~GHz & %
  0.27 & 0.20 & \nodata \\
\spider 2 & $2 \times 90$~GHz; $2 \times 150$~GHz; $2 \times 280$~GHz & %
  0.21 & 0.16 & 0.62
\enddata
\tablecomments{The multiple telescopes at each frequency are rotated
  with respect to one another to simultaneously recover $Q$ and $U$
  in each daily map.  The distribution of observing frequencies is
  chosen following the procedure described in
  Section~\ref{sec:bs+fr} so as to maximize \spidern's
  ability to detect inflationary $B$-modes at angular scales
  corresponding to $\ell\sim 80$.  The cumulative noise
  numbers use the parameters listed in Table~\ref{tbl:detectors}.}
\end{deluxetable}

\begin{figure}[t]
\centering
\includegraphics[height=2.35in]{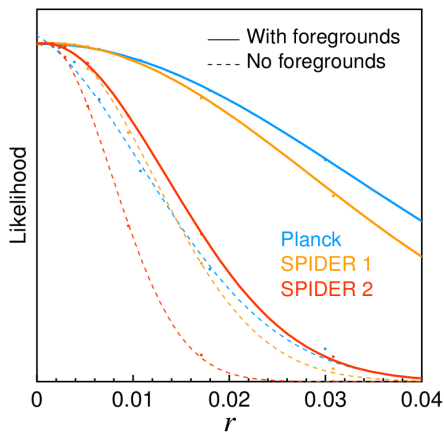}
\caption{Marginalized one-dimensional $r$-likelihood curves
  for \spider and \Planck~HFI.  In the ``with foreground'' case,
  foreground-subtraction residuals are included as contaminants.  The
  ``no foreground'' case assumes no foreground contamination.
  In the absence of foregrounds,
  the first flight of \spider (\spider~1) achieves the $3\sigma$ limit
  $r < 0.03$, while the two flights combined (\spider 2) lead to 
  $r < 0.02$ (99\%~CL).  In the ``with foreground'' case, two flights are
  needed to reach the $3\sigma$ limit $r<0.03$.  The details of the
  two flights are given in Table~\ref{tab:fpuflights}. 
  \label{fig:inflation_lims1}}
\end{figure}

Figure~\ref{fig:inflation_lims1} shows forecasts for the constraints
on the tensor-to-scalar ratio $r$ from \spider (after one and both
flights) and from \Planck.  These constraints are 
derived using the method described in detail in
Farhang et al.~(submitted), which we summarize~below. 

Assuming a uniform {\it a priori} probability distribution for the
cosmological parameters $\mathbf{q}$, the {\it a~posteriori}
distribution function's dependence on $\mathbf{q}$ is given by the
likelihood
${\cal L}_\mathbf{\Delta} (\mathbf{q}) \equiv P(\mathbf{\Delta}|\mathbf{q})$, 
the probability of the data $\mathbf{\Delta}$ given the parameters,
such that
\begin{equation}
\label{eqn:likelihood}
-2 \ln {\cal L}_\mathbf{\Delta} (\mathbf{q}) =
\mathbf{\Delta}^\dagger\,\mathbf{C}^{-1}_{\rm tot}\,\mathbf{\Delta} +
\ln |(2\pi)^{N_{\rm pix}}\, \mathbf{C}_{\rm tot}|,
\end{equation}
where $\mathbf{\Delta}$ is the full set of CMB temperature and
polarization maps including both signal and noise,
$N_{\rm pix}$ is the number of pixels in the maps, and
$\mathbf{C}_{\rm tot} \equiv \mathbf{C}_{\rm N} + %
 \mathbf{C}_T (\mathbf{q})$
is the theoretical pixel-pixel covariance matrix with 
contributions from both the parameter-dependent signal covariance
$\mathbf{C}_T (\mathbf{q})$ and the generalized noise 
covariance~$\mathbf{C}_{\rm N}$.  The latter includes uncertainties
from foreground subtraction as well as from noise in the maps.
 
Here, we calculate the likelihood in Equation (\ref{eqn:likelihood})
on a two-dimensional grid consisting of the cosmological
parameters $r$ and $\tau$, the latter referring to the Thomson
scattering optical depth to reionization.  Although extra dimensions
could be added, Farhang et al.~(submitted) show that $r$ is
only weakly correlated with other standard cosmological
parameters in the small-$r$, small-$f_\sky$ limit of relevance to \spidern.
Since stringent constraints on $\tau$ are an objective of 
large-scale experiments such as \Planck, we marginalize over it in
order to evaluate whether insensitivity to it will impact \spidern's
ability to constrain $r$.  It appears that with our sky coverage fixed
at 10\%, $r$ and $\tau$ are essentially uncorrelated.

The results in Figure~\ref{fig:inflation_lims1} assume $f_\sky=0.75$
for \Planck~and $f_\sky=0.10$ for \spidern, and are given for two
cases: ``with'' and ``without'' foregrounds.  In the former case, 
foreground residuals after the removal procedure described earlier in
this section are included as contaminants, whereas in the latter case,
it is assumed that there is no foreground contamination in the maps.
Since we seek to derive the most stringent upper limit on $r$
that \spider could set given the assumptions in this paper,
the input cosmological model assumes $r=0.001$ (and $\tau=0.09$).
Farhang et al.~(submitted) address the question of how well
``large''-$r$ 
$B$-modes can be detected.

We find that, with two flights, \spider has the ability to constrain
the value of the tensor-to-scalar ratio $r$ to below $0.03$~(99\%~CL).
This number improves to $0.02$~(99\%~CL) in the limit of no foreground
contamination.  It is worth noting that the constraint quoted above
for the ``with foreground'' case is based on the foreground removal
technique described earlier in this section, which assumes that the
frequency dependence of the polarized dust emission is well described
by a single power law.  Although this is in line with the current
state-of-the-art for $B$-mode forecasts
\citep[see, e.g.,][]{Dunkley_etal_2009}, the detailed question of how
deviations from this description will be handled has yet to be
studied.  Any such deviation will likely impact the quoted constraints
\citep[see, e.g.,][]{Armitage-Caplan_etal_2012}.  Relatedly, it is
possible that a different frequency dependence of the polarized dust
emission might require a different frequency coverage in order to
reach the best possible constraints on $r$.  Should this become
apparent after the first \spider flight, or through improvements in
our knowledge of foregrounds before \spider flies, the flexibility of
\spidern's focal plane composition discussed in
Section~\ref{sec:inst_description} ensures that we will be able to
adopt the best frequency coverage to reach our science goals.


\section{Summary \& conclusions}
\label{sec:final}

Our main results are as follows.
\begin{enumerate}
\item The \spider instrument was designed with an emphasis on
  reaching low levels of polarized systematic error at multipoles
  close to the $\ell\sim 80$ acoustic peak in the $B$-mode power
  spectrum (Section~\ref{sec:inst_description}).  Taking into account
  all known sources of systematic error, the residual power spectrum
  is less than 37\% of the $B$-mode signal for $r=0.03$ at $\ell=100$
  (Section~\ref{sec:systematics}).  As we refine our systematics
  budget, we expect this number to go down
  for three reasons: (1)~more stringent design targets than those
  shown in Table~\ref{tbl:Systematics} have already
  been reached; (2)~recent updates to our flight plan and to our
  scanning strategy are expected to reduce the overall level of
  systematic error by providing better observing conditions; and
  (3)~all residuals in Table~\ref{tbl:Systematics} were added in 
  quadrature, which leads to a conservative estimate of the global
  systematic error.  Simulations taking these three points into
  account are underway.  The reduction in systematic error that we
  expect these improved simulations to show will likely be partially
  canceled by the inclusion of systematic effects that are yet to be
  characterized, in particular in so far as they relate to thermal
  stability of the focal plane and of the optical sleeve.
\item Polarized sidelobes should not hinder \spidern's ability to
  detect or constrain $B$-modes.  Using the polarized sidelobe profile 
  measured in \bicepn, we showed that the corresponding contamination at
  $\ell = 100$ is over an order of magnitude fainter than the $r=0.03$ 
  $B$-mode signal (Section~\ref{sec:polsidelobes}).  As illustrated 
  in Figure~\ref{fig:polsidelobe}, improved baffling will lower this
  number by over an order of magnitude.  At the largest
  scales ($\ell\sim 10$), it will lead to a polarized
  sidelobe signal of amplitude comparable to that of the $r=0.03$
  $B$-modes.
\item \spidern's observing strategy (Section~\ref{sec:obs_overview})
  was designed to minimize the $T$/$Q$/$U$ covariance,
  thereby minimizing the uncertainty in the reconstruction of the
  Stokes parameters.  Figure~\ref{fig:crosslink} shows that, taking
  all detectors into account, the excess variance introduced by
  the scanning strategy is less than $1\%$ over $96\%$ of the sky
  observed during a 4-day~period.
\item The lack of atmospheric contamination at stratospheric altitude
  and large sky coverage enabled by \spidern's small apertures
  provide a relatively broad dynamic range in multipole, as shown by
  the filter transfer function ($F_\ell$) study in
  Section~\ref{sec:fell}.  This opens the possibility of moving beyond
  detection to a spectral characterization of the $B$-mode power
  spectrum, an important consistency check on the cosmological origin
  of any detected signal.  Furthermore, the fidelity of the
  measurement results in a significantly higher final sensitivity than
  would be achieved by an experiment of the same NET using a less
  efficient observation platform.
\item The frequency distribution as a function of \spider flight
  summarized in Table~\ref{tab:fpuflights} ensures high
  signal-to-noise measurements of the microwave sky at low-foreground 
  frequencies, while the polarized emission from interstellar dust is 
  mapped with high sensitivity in a band where it overwhelmingly
  dominates.  With this distribution, we showed in
  Section~\ref{sec:bs+fr} that \spider will
  provide the stringent constraint $r<0.03$ (99\% CL) after both
  flights, even when accounting for residual foreground
  contamination.  This number is derived under the assumption that the  
  \Planck~217~GHz data will be available during the \spider analysis.
  When foreground contamination is ignored, \spidern's
  constraint improves to $r<0.02$ (99\% CL).
\item Among Galactic foregrounds, we expect the polarized emission
  from interstellar dust to be the only signal as bright or brighter
  than the $r=0.03$ $B$-modes at \spider frequencies
  (Section~\ref{sec:spider_foregrounds}).  We model this emission
  following the procedure detailed in Section~\ref{sec:dust_model},
  and show that the \spider observing region includes the cleanest
  $2\%$ of the sky accessible from a McMurdo LDB flight.  The average
  amplitude of the polarized Galactic dust emission in the \spider
  field is comparable to that in a typical $f_\sky=2\%$ patch in the
  ``Southern Hole;'' many $f_\sky=2\%$ patches in the \spider region
  exhibit significantly less polarized dust emission.  This study
  suggests that without component separation,
  degree-scale polarized dust emission will limit the constraints of
  {\it any} experiment at or above the level of $r\sim0.03$, {\it even
    in the portions of the southern sky most free of
    Galactic dust emission}.
\end{enumerate}

Although more work is needed to fully characterize the \spider
instrument and its nontrivial interaction with the microwave sky, a
simple Galactic foreground situation could allow \spider to
characterize a cosmological $B$-mode signal whose intensity is close
to the lowest compatible with the simplest viable class of
inflationary models.  General studies of the impact of a more complex
foreground situation have yet to be performed, and
are of obvious interest in the context of~\spidern.


\acknowledgments 

The \spider collaboration gratefully acknowledges the support of the
National Science Foundation (ANT-1043515), NASA
(APRA-NNX07AL64G), and
the Gordon and Betty Moore Foundation.  Support in Canada is provided
by NSERC, the Canadian Space Agency, and CIFAR.  JEG is supported by a
grant from the Leifur Eir\'iksson Foundation.  ASR is supported by
NASA (NESSF-NNX10AM55H).  WCJ acknowledges the generous support of the
Alfred P. Sloan Foundation and of the David and Lucile Packard
Foundation.  Some of the results in this paper have been derived using
the HEALPix\footnote{\texttt{http://healpix.jpl.nasa.gov}}
\citep{Gorski_etal_2005} package, as well as the FFTW subroutine
library \citep{Frigo+Johnson_2005}.  This research has made use of
NASA's Astrophysics Data System. We acknowledge the use of the Legacy
Archive for Microwave Background Data Analysis (LAMBDA). Support for
LAMBDA is provided by the NASA Office of Space Science.


\bibliographystyle{JHEP}
\bibliography{refs_rev2}

\end{document}